\documentclass[journal]{IEEEtran}
\ifCLASSOPTIONcompsoc
  \usepackage[nocompress]{cite}
\else
  \usepackage{cite}
\fi
\ifCLASSINFOpdf
\else
\fi
\usepackage{graphicx}
\usepackage{amsmath}
\usepackage{subfigure}
\usepackage{multirow}
\usepackage{graphicx}
\usepackage{amsmath}
\usepackage{multirow}
\usepackage{amsfonts,amsmath}
\usepackage{amssymb, nccmath}
\usepackage{amssymb, nccmath}
\usepackage{dsfont}
\usepackage{makecell}
\usepackage{booktabs}
\usepackage[utf8]{inputenc}
\usepackage{dsfont}
\usepackage{booktabs}
\usepackage{graphicx}
\usepackage{grffile}
\usepackage{boldline}
\usepackage{grffile}
\usepackage{array} 
\usepackage{mathtools}
\usepackage{fontenc}
\usepackage{tikz}
\usepackage{supertabular,enumitem}
\usepackage{tikz}

\usepackage{pgfplots}

\definecolor{mygreen}{HTML}{03C03C}
\definecolor{mypink}{HTML}{008000}
\definecolor{Mycolor}{HTML}{FFC40C}
\usepackage[linesnumbered, ruled]{algorithm2e}
\makeatletter
\newcommand{\removelatexerror}{\let\@latex@error\@gobble}
\makeatother
\usepackage{subfig}
\usepackage{fancyhdr}
\renewcommand\thesection{\arabic{section}}

\renewcommand{\thesubsubsection}{\Alph{subsubsection}}
\begin{document}
\title{A Secure and Multi-objective Virtual Machine Placement Framework for Cloud Data Centre}
\author{Deepika~Saxena,~Ishu~Gupta,~\textit{Member, IEEE,}~Jitendra~Kumar,~Ashutosh~Kumar~Singh,~\textit{Senior Member, IEEE,}~and~Xiaoqing~Wen,~\textit{Fellow, IEEE}
	%and~Jane~Doe,~\IEEEmembership{Life~Fellow,~IEEE}% <-this % stops a space
	\IEEEcompsocitemizethanks{\IEEEcompsocthanksitem D. Saxena, I. Gupta and A. K. Singh are with the Department of Computer Applications, National Institute of Technology, Kurukshetra, India. J. Kumar is with Department of Computer Applications, NIT Tiruchirappalli, Tamilnadu, India. E-mail: 13deepikasaxena@gmail.com, ishugupta23@gmail.com, jitendra@nitt.edu, ashutosh@nitkkr.ac.in \\	
		X. Wen is with the Department of Creative Informatics and the
		Graduate School of Computer Science and Systems Engineering, Kyushu
		Institute of Technology, Fukuoka 8208502, Japan (e-mail:
		wen@cse.kyutech.ac.jp)
}}

\IEEEtitleabstractindextext{%
\begin{abstract}
To facilitate cost-effective and elastic computing benefits to the cloud users, the energy-efficient and secure allocation of virtual machines (VMs) plays a significant role at the data centre. The inefficient VM Placement (VMP) and sharing of common physical machines among multiple users leads to resource wastage, excessive power consumption, increased  inter-communication cost and security breaches. To address the aforementioned challenges, a novel secure and multi-objective
virtual machine placement (SM-VMP) framework is
proposed with an efficient VM migration. The proposed framework ensures an
energy-efficient distribution of physical resources among VMs that emphasizes
secure and timely execution of user application by reducing
inter-communication delay. The VMP is carried out by applying the proposed Whale Optimization Genetic Algorithm (WOGA), inspired by whale evolutionary optimization and non-dominated sorting based genetic algorithms. The performance evaluation for static and dynamic VMP and comparison with recent state-of-the-arts observed a notable reduction in shared servers, inter-communication cost, power consumption and execution time up to 28.81\%, 25.7\%, 35.9\% and 82.21\%, respectively and increased resource utilization up to 30.21\%.

\end{abstract}

}

\maketitle
\thispagestyle{fancy}

\IEEEdisplaynontitleabstractindextext

\renewcommand\thesubsection{\Alph{subsection}}

% \fi
\IEEEpeerreviewmaketitle

{\section{Introduction}\label{sec:introduction}}
\IEEEPARstart{T}{he}
 minimum upfront capital investment and maximum scalability features of cloud computing are fascinating to businesses, academia,  research and all public or private organizations \cite{sharma2016multi}, \cite{kumar2020biphase}, \cite{saxena2018abstract}. The smooth and progressive working of these organizations depend on the services offered by cloud data centre \cite{singh2021quantum}, \cite{saxena2016dynamic}. To meet the ever-growing and dynamic demand of users, more and more VMs are deployed on a large number of servers and cooling devices are installed at data centre that account for high power consumption \cite{saxena2021op}, \cite{saxena2020auto}, \cite{saxena2015highly}. The VM placement (VMP) has a crucial impact on resource utilization, power consumption, and the overall operational cost of the data centre \cite{kumar2021resource}, \cite{saxena2021workload}, \cite{saxena2015ewsa}. 
The inefficient VMP leads to under/over-utilized servers with increased power consumption and resource wastage. On the other hand, the VMP with efficient resource utilization allows multiplexing and sharing of physical resources, introduces risk of security breaches such as side-channel attack that may cause VMs sharing common physical machine (co-residency) to steal sensitive information \cite{han2015using}, \cite{gupta2019confidentiality}, \cite{gupta2017probability}. Generally, users are completely unaware about the actual location of their VMs and run their confidential applications ignoring the probability of security attacks \cite{gupta2020mlpam}, \cite{chhabra2020secure}, \cite{singh2021cryptography}. Infact, co-residency raises occurrence of security threats by giving extensive opportunity to the attacker to exploit vulnerabilities of hypervisor and compromise other user's VM \cite{saxena2021oscmc}. Hence, it becomes the responsibility of the cloud resource manager to mitigate the vulnerability of user's application by considering security as an important objective and intuitively minimize co-location of different user's VM on common server during VMP \cite{saxena2020security}. 
Moreover, VMs access physical network to communicate with interdependent VMs located at different servers to execute data-intensive applications, which hampers cloud client’s experience, raises network traffic and degrades overall performance by saturating the network links \cite{luo2018reliable}. Intentionally, VMs executing common application must be placed in closer proximity to reduce inter-communication cost within the data centre \cite{jennings2015resource}. Therefore, the aforementioned entangled issues generate a complex and challenging multi-objective VMP problem, in which resource capacity constraints: CPU, RAM, storage and bandwidth etc. must be satisfied, before actual placement of VMs on servers. Such a VMP problem belongs to NP-Complete complexity class \cite{duong2018dynamic}.

The minimum power consumption and maximum resource utilization are utmost concerns to cloud service provider to reduce overall operational cost while security and response-time are the major perspectives for cloud users. Considerable research works are available to handle VM consolidation issues with cloud service providers' perspective that focussed on energy-efficient resource utilization. For instance, the energy-efficient VMP approaches based on heuristic/meta-heuristic optimization is presented in \cite{saxena2021energy} in which over/under-resource utilization is tackled with energy proficient VMP so that the entire workload gets consolidated on the minimum number of active servers \cite{chhabra2018probabilistic}, \cite{kumar2021self}. Also, some existing research proposals have considered security and communication cost of data-intensive applications, for example, secure VMP by minimizing the risks of side-channel attacks, is discussed in \cite{han2015using}, \cite{kumar2018workload}. The network traffic aware VMP is explored in \cite{saxena2021carelb}, in which network bandwidth and traffic are optimized by allocating inter-dependent VMs on closer placed servers. 
 \par To serve the quest of incorporating interests of both cloud user and service provider in conjunction, it is essential for VMP fraternity to give unanimous importance to each objective. This demands a multi-efficient VMP framework using the dominant paradigm of multiplexing and physical resource sharing. This work presents a \textbf{S}ecure and \textbf{M}ulti-objective \textbf{V}irtual \textbf{M}achine \textbf{P}lacement (SM-VMP) framework to address the challenges above. A novel WOGA multi-objective VMP optimization algorithm is proposed using concepts of {W}hale {O}ptimization algorithm \cite{mirjalili2016whale} and non-dominated sorting based {G}enetic {A}lgorithm \cite{deb2002fast} to search the most optimal VMP subject to security, resource utilization, communication-cost and power consumption.

  \textit{Key Contributions}:
\begin{itemize}
	\item 
	
	SM-VMP framework is presented for placement of user's VM subject to minimum security threats, communication cost, power consumption and improved resource utilization. To the best of the authors knowledge, these four objectives have been addressed concurrently for the first time during resource allocation.
	\item A novel WOGA multi-objective VMP optimization algorithm is proposed to achieve the feasible and optimal VMP to serve both cloud users and service providers' perspectives unanimously.  
	
	\item A probability based encoding-decoding system is introduced, which encodes VM placement as a whale position vector by considering the heterogeneous VM environment with respect to the number and type of VMs between two VM allocation vectors.
	
	\item Multi-efficient VM migration is proposed for dynamic allocation/de-allocation of VMs.
 
\end{itemize}

%\subsection{Organization}
 \textit{Organization}: Section 2 discusses the related work. Section 3 entails System Model followed by SM-VMP framework and multi-efficient VM migration are presented in Section 4 and Section 5, respectively. The performance evaluation including experimental set-up, results and comparison with state-of-the-arts for static and dynamic VMP, is presented in Section 6 followed by conclusive remarks and future scope of the proposed work in Section 7. Table \ref{table:notation} shows the list of symbols with their explanatory terms used in this paper. 
\begin{table}[htbp]
	\centering
	
	\caption[Table caption text] {Notations and their descriptions}  %\cite[p.10]{refid} }
	\label{table:notation}
	\resizebox{1.0\textwidth}{!}{\begin{minipage}{\textwidth}
			\begin{tabular}{l}
				\hline
				%\multicolumn{2}{c}{Item} \\
				%\cline{1-2}
				%Symbols: Explaination terms  \\				\hline
				{$\hat{D}$}{: distance between best/random and current solution;}\\  {$\hat{C}$, $\hat{A}$: Co-efficient vectors;}
				%$\hat{C}, \hat{A}$          & co-efficient vectors      \\
				{$\psi_n$: $n^{th}$ VM allocation;}\\   {$x$: variable decreasing from 2 to 0 ; 
				 $\hat{r}$: random value [0,1];}\\ {$S_i$: $i^{th}$ server;}      
			
				 {$v_j$: $j^{th}$ VM;} {$\omega_{ji}$: Mapping of $j^{th}$ VM on $i^{th}$ server;}\\
				
				{$RU$: Resource utilization;  $PW$: Power consumption;} \\
			
				{$\phi$: No. of conflicting servers; $\vartheta$: Communication cost;} \\
		
				{$\gamma$: Status of server; $N$: Number of resources}\\
				{$\alpha_{ki}$: $i^{th}$ server has VM of $k^{th}$ user;} {$ob$: Number of objectives;}\\
				{$\beta{ik}$: $k^{th}$ user has VM on $i^{th}$ server; $cp$: Cross-over point;}\\ 
				%$ob$: Number of objectives; $cp$: Cross-over point\\  
				{$\eta$: Cost function; $X$: Number of VM allocations/solutions}\\
			
				%$G_{max}$& maximum number of iterations\\
				{$C_1$, $C_2$, $C$: Offspring produced after crossover;} \\
			{$\mu(C_1)$, $\mu(C_2)$: Mutation operation on offsprings after crossover}\\
				\hline
			\end{tabular}
	\end{minipage}}
\end{table}

\section{Related Work}

The pioneering works presented for population based VMP approaches includes Genetic Algorithm (GA), swarm intelligence such as Particle Swarm Optimization (PSO), Ant Colony Optimization (ACO) and whale optimization (WOA) \cite{donyagard2019multiobjective}. The GA based approaches are applied in numerous research works including \cite{singh2019secure},  \cite{tseng2017dynamic}, and \cite{saxena2020proactive} etc. 
Multi-objective GA for resource prediction and allocation is proposed in \cite{tseng2017dynamic}, that predicts resource usage before VM allocation to maximize resource utilization and energy saving. A GA and weighted sum approach based optimized VMP in \cite{tseng2017dynamic}, where GA is applied to predict the resource utilization of VMs followed by their energy-efficient allocation on servers. Recently, Singh et al. \cite{singh2019secure} presented NSGA-II based approach and introduced the security concept during VMP, by minimizing the number of conflicting servers along with power saving and efficient resource utilization. The limitation of GA based VMP is that it often leads to premature convergence. A Security-aware Multi-Objective Optimization based
  virtual machine Placement (SMOOP) is proposed in \cite{han2017reducing} by applying a modified GA and weighted sum approach to generate optimal VMP subject to security risks, network traffic and resource utilization but, ignored the role of power consumption. {Al-Dulaimy et al. \cite{al2018type} have classified the VMs according to the type of applications to enhance the energy consumption, and further, on top of this work, they built security and privacy restrictions for the VM placement process in \cite{al2019privacy}.}   
  Sharma et al. \cite{sharma2016multi} presented a hybrid PSO and GA i.e., HGAPSO algorithm to minimize resource wastage and SLA violation during VM allocation. This approach encodes VM allocation as particle vector where a bit value is 1 if the server is active, otherwise, it is 0. This method is suitable for homogeneous VMP because the bit value of velocity vector depends on presence or absence of VMs only and do not perfectly encode the number and type of VMs, hence not suitable for heterogeneous environment. In addition, energy and resource efficient VM allocations based on whale optimization and its hybrid approaches are presented in \cite{strumberger2019resource}, \cite{rana2018cloud}. Basset et al. proposed a bandwidth-efficient VMP in \cite{abdel2019improved} by applying improved levy based WOA hybridized with best-fit strategy. These works considered homogeneous configuration of VMs with static allocation of physical resources. The random-fit, best-fit and first-fit heuristic for VMP are discussed in \cite{jung2010mistral}, \cite{shirvastava2017best} and \cite{jangiti2019aggregated} respectively. However, these heuristic approaches do not provide best suited optimal solution for multi-objective problem and best-fit algorithm is not scalable in nature due to long convergence time.
\par {Unlike existing VMP approaches, SM-VMP serves interests of both service provider as well as user by optimizing resource utilization, security, networking cost and power consumption, concurrently, during VM allocation. Moreover, the existing works applied same approach for homogeneous as well as heterogeneous VMs. The heterogeneous configuration of different VMs and servers is not taken into account, which makes a big difference for large scale data centres. The proposed WOGA optimization introduced probability values based encoding-decoding system, which encodes VM placement subject to their heterogeneous configuration to generate optimal VM placement that not only ensures energy-efficient resource utilization but also emphasizes security and timely execution of user applications.}

\section{System Model}
\begin{figure}[htbp]
	\centering
	\includegraphics[width=0.8\linewidth]{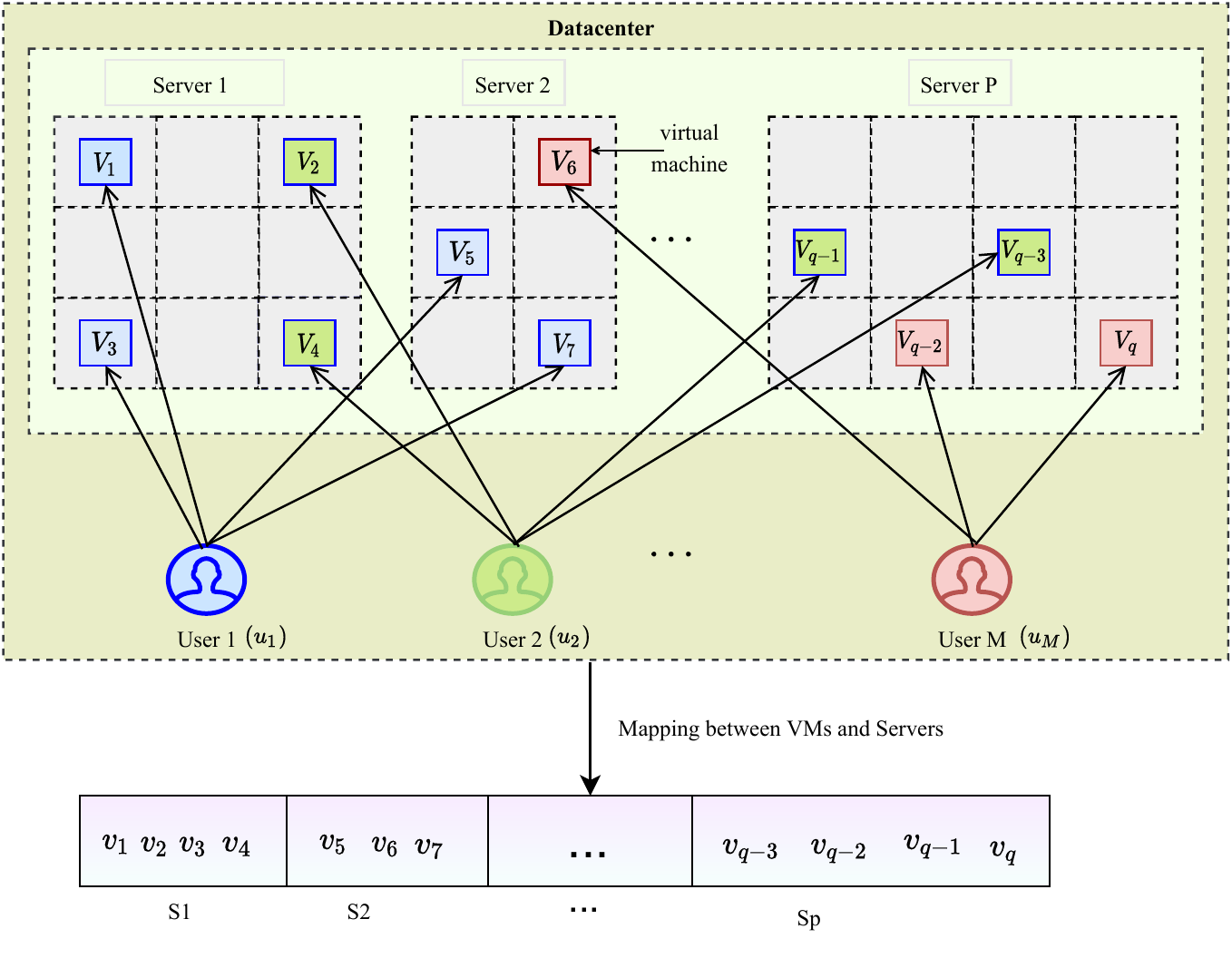}
	\caption{ VM allocation at Cloud data centre}
	\label{scenario}
\end{figure}

Consider a data centre having $P$ servers as $ {S_1, S_2,...,S_p} \in S$ and $Q$ virtual machines (VMs) as $ {v_1,v_2,...,v_q} \in V$, both having heterogeneous configuration as shown in Fig. \ref{scenario}. There are $M$ users such that ${u_1,u_2,...,u_M} \in U$ who have purchased different VMs at the data centre where the physical allocation of VMs is shown by mapping between VMs and servers. The filled and vacant blocks represent allocated and de-allocated/unoccupied VMs respectively on a server. 
 The proposed SM-VMP framework optimizes resource distribution among VMs to cater cloud user as well as service-provider's perspectives by promising maximization of resource utilization, and minimization of security breaches, communication-cost and power consumption. Four distinct models associated to each objective are designed and trade-off among them is discussed as follows:

	\subsection{Resource utilization} 
Let $S_i^{C}$, $S_i^{Mem}$ and $S_i^{RAM}$ are CPU, memory, and RAM capacity of $i^{th}$ server. Similarly, for $j^{th}$ VM, CPU, memory, and RAM utilization are represented as $v_j^{C}$, $v_j^{Mem}$ and $v_j^{RAM}$ respectively. If $S_i$ is in operational mode then $\gamma_i=1$, otherwise $0$. Let $\omega$ is a two-dimensional matrix of size $P \times Q$ which represents mapping of VMs to servers having vector size $Q$ and $P$ respectively. If server $S_i$ hosts $v_j$, then $\omega_{ji}=1$ otherwise $0$. The usage of each resource is separately monitored by the model where, CPU, memory and RAM utilization of a server can be computed by applying Eqs. (\ref{cpu})-(\ref{ram}). 
	\begin{equation}
	RU_i^{C}=\frac{\sum_{j=1}^{Q}{\omega_{ji} \times v_j^{C}}}{S_i^{C}}\label{cpu}
	\end{equation}
	\begin{equation}
	{RU_i}^{Mem}=\frac{\sum_{j=1}^{Q}{\omega_{ji} \times v_j^{Mem}}}{S_i^{Mem}}\label{memory}
	\end{equation}
	\begin{equation}
	{RU_i}^{RAM}=\frac{\sum_{j=1}^{Q}{\omega_{ji} \times v_j^{RAM}}}{S_i^{RAM}}\label{ram}
	\end{equation}
	 Eq. (\ref{cpu1}) evaluates resources utilization of data centre ($RU_{dc}^{\mathds{R}}$) and complete resource utilization ($RU_{dc}$) is determined by applying Eq. (\ref{ru}) where, {$N$ is the number of resources}. 

	\begin{equation}
	RU_{dc}^{\mathds{R}}= \sum_{i=1}^{P}{RU_i^{\mathds{R}}} \quad  \mathds{R} \in \{C, Mem, RAM\} \label{cpu1}
	\end{equation}
	
	\begin{equation}
	RU_{dc}= \int\limits_{\substack{t_1\\\mathcal{}}}^{t_2} \bigg(\frac{	RU_{dc}^{C} +  RU_{dc}^{Mem} + RU_{dc}^{RAM} }{|N|\times \sum_{i=1}^{P}{\gamma_i}} \bigg){dt}\label{ru}
	\end{equation}
	\subsection{Security} 
	In order to reduce the security breaches, the allocation of VMs of different users on the same server is minimized to combat the chances of security attack through co-residency. This will reduce the probability of occurrence of side-channel attacks ($\phi$), where a malicious VM targets the co-resident VMs for stealing confidential information. {Let $\alpha_{ki}$ be the mapping between user $u_k$ and server $S_i$, where $\alpha_{ki}=1$, if a server hosts one or more VMs of user $u_k$, otherwise $\alpha_{ki}=0$. The number of users having their VMs allocated on a server $S_i$ can be calculated as $\sum_{k=1}^{M}{\alpha_{ki}}$. The presence of the conflicting servers' percentage can be obtained from Eq.~\eqref{security}}. As opposed to the existing secure virtual machine assignment schemes such as~\cite{han2017reducing}, the proposed approach ensures the reduction of co-residential vulnerability threats without any prior information of the presence of a malicious VM.
	\begin{equation}
	{\phi_{dc}}= \int\limits_{\substack{t_1\\\mathcal{}}}^{t_2}\bigg(\frac{\sum_{i=1}^{P}\sum_{k=1}^{M}{\alpha_{ki}}}{|S|} \bigg){dt} \times 100 ;   \quad \forall \sum_{k=1}^{M}{\alpha_{ki} > 1}
	\label{security}
	\end{equation} 

	%\item \textit
	\subsection{Communication cost}
	Assume that VMs of a common user are inter-dependent (i.e., communication intensive), {servers are packed into cluster in linear arrangement and distance between any two consecutive servers within a cluster is counted as `one hop'}. Let $\beta_{ik}=1$ denotes one or more VM(s) of user $u_k$ is deployed on server $S_i$ otherwise, it is 0. The distance between servers hosting VMs of a common user denoted as $\hat{\lambda}$, is computed by counting number of hops by applying Eq. (\ref{diff}), where $S_i^{loc}$ and $S_j^{loc}$ are location of servers $S_i$ and $S_j$, respectively. The number of users having their VMs on farther placed servers, are represented as $\Xi$, which can be computed depending upon the value of $\hat{\lambda}$ as stated in Eq. (\ref{count}). The counter $\Xi$ is incremented if $\hat{\lambda} > c$, where $c$ is any constant number to define limits of distance between closer placed servers. For experiment, {we increase} $\Xi$ if $\hat{\lambda} \ge 2$, which means if VMs of a user are hosted on more than two consecutive servers in a cluster, then they are considered to be placed, farther, though these settings can be changed as per the requirement. Eq. (\ref{comm}) determines percentage of such user at an instant \{$t_1, t_2$\}, which is to be minimized.  

	\begin{equation}
	{\hat{\lambda_k} = |S_i^{loc} \times \beta_{ik} - S_j^{loc} \times\beta_{jk}|}; \forall k \in \{1, M\}; i,j \in \{1, P\}, i\neq j	\label{diff}
	\end{equation}
	\begin{gather}\label{count}
		\Xi =\Xi +1; \quad \forall_k \hat{\lambda_k} > c, k \in {1, M} 
	\end{gather}
	
	\begin{gather}
		\vartheta_{dc}= \int\limits_{\substack{t_1\\\mathcal{}}}^{t_2}\frac{\Xi}{|U|}{dt} \times 100
	\label{comm}
	\end{gather}       
The communication cost is minimized by allocating VMs of single user at minimum number of servers in close proximity. This would reduce the VM's dependency on physical network access for execution of an application and therefore, communication cost within the data centre and response time of data-intensive application is also reduced.
	      
	\subsection{Power consumption}
	The ample amount of high power consumption by servers, accounts for high operational cost at data centre that needs to be minimized. The power consumption is contributed by CPU, memory, storage and number of active network cards in operational mode. The proposed work considers all the servers based on inbuilt Dynamic Voltage Frequency Scaling (DVFS) energy saving technique \cite{minas2009energy} based on the two states of CPU: sleep and busy state. In sleep mode, CPU works in minimum processing mode with reduced clock cycle and some internal components of CPU are in non-operational mode. During busy state, power consumption depends on the processing application and CPU utilization rate. Therefore, power consumption for a server can be formulated as $PW_i$ for $i^{th}$ server and total power consumption $PW_{dc}$ for time-interval \{$t_1$, $t_2$\} as shown in Eqs. (\ref{power1}) and (\ref{power2}), respectively, where $RU_{i} \in$ {[0, 1]} is $i^{th}$ server’s resource utilization.

	\begin{equation}
	PW_i =([{PW_i}^{max} - {PW_i}^{min}] \times RU_{i} + {PW_i}^{idle}) \label{power1}
	\end{equation} 
	
	\begin{equation}
	PW_{dc} = 
	\int\limits_{\substack{t_1\\\mathcal{}}}^{t_2} \bigg(\sum_{i=1}^{P} {PW_i}\bigg){dt}
	\label{power2}
	\end{equation}	
%\end{itemize}
\subsection{Trade-off among multiple objectives}

The resource utilization and reduction of power consumption tend to improve with consolidation of VMs on minimum number of active servers while security and network communication cost are directly affected by number of active users at an instant. If a user has more VMs unable to accommodate on a single server, probability of security threats and inter-communication cost increase. %As the number of users scales up, there arises risk of threats. 
Obviously, security increases at the cost of resource utilization and power consumption. However, placing inter-dependent VMs in closer physical location results into reduced VM migration, decreased network traffic as well as power consumption, since both are strongly correlated \cite{baldoni2014correlating}. When power consumption is small, the entire load is distributed on small number of servers, that may cause resource contention among VMs and degrade performance.

 \subsection{Constraints} The essential set of VMP constraints that must be satisfied has been formulated as given in Eqs. (\ref{1})-(\ref{4}):
 
\begin{equation}
C_1: \quad \sum_{j=1}^{Q}{v_j^{pe}} \times \omega_{ji} \leq S_i^{pe} \quad  \forall i \in {1,2,...,P}\label{1}
\end{equation}

\begin{equation}
C_2:\quad  \sum_{j=1}^{Q}{v_j^{C}} \times \omega_{ji} \leq S_i^{C} \quad  \forall i \in {1,2,...,P}\label{2}
\end{equation}

\begin{equation}
C_3:\quad  \sum_{j=1}^{Q}{v_j^{RAM}} \times \omega_{ji} \leq S_i^{RAM} \quad  \forall i \in {1,2,...,P}\label{3}
\end{equation}
\begin{equation}
C_4: \quad \sum_{j=1}^{Q}{v_j^{Mem}} \times \omega_{ji} \leq S_i^{Mem} \quad  \forall i \in {1,2,...,P}\label{4}
\end{equation}
where $\omega_{ji}$ represents mapping of $j^{th}$ VM on $i^{th}$ server, $v_j^{pe}$ and $S_i^{pe}$, $v_j^{C}$ and $S_i^{C}$, $v_j^{RAM}$ and $S_i^{RAM}$, and $v_j^{Mem}$ and $S_i^{Mem}$ are number of processing element, capacity 
of CPU (in MIPS), RAM and secondary storage requirements of $j^{th}$ VM and available for $i^{th}$ server, respectively. The above discussion implies that considered problem is a specialized case of multi-objective VMP optimization entangled with contradictory challenges and requires a solution to allow energy-efficient physical resource utilization with secure and timely execution of user application. This problem is formulated using Eqs. (\ref{vm_allocation}) and (\ref{model}) where $\psi$ represents allocation of $Q$ VMs to $P$ servers such that objectives: $PW$, $\vartheta$, $\phi$ are minimized and $RU$ is maximized subject to constraints specified in Eqs. (\ref{1})-(\ref{4}).
\begin{equation}
\label{vm_allocation}
{	\psi=\sum_{i=j=1}^{P,Q}{\omega_{ji}}}
\end{equation}
\begin{equation}\label{model}
\centering
%\resizebox{0.49\textwidth}{!}{$
	\begin{aligned}	
{Minimize\quad f_{PW_{dc}}\big(\psi \big), f_{\vartheta_{dc}}\big(\psi \big), f_{\phi_{dc}}\big(\psi \big)},\\ {f_{-RU_{dc}}\big(\psi \big) \quad s.t. \quad \{C_1, C_2, C_3, C_4\}}
\end{aligned}
\end{equation}

%\end{align} 
To resolve aforementioned problem, WOGA VMP evolutionary optimization algorithm is developed which integrates exploration and exploitation capabilities of WOA (which is computationally efficient algorithm having fewer parameters with effective properties to explore the best solution for optimization problems \cite{mehne2018parallel}) and pareto-optimal solution guided by non-dominated sorting based GA.

\section{ SM-VMP Framework} 
\label{SM-VMP}
\begin{figure}[htbp]
	\centering
	\includegraphics[width=0.85\linewidth]{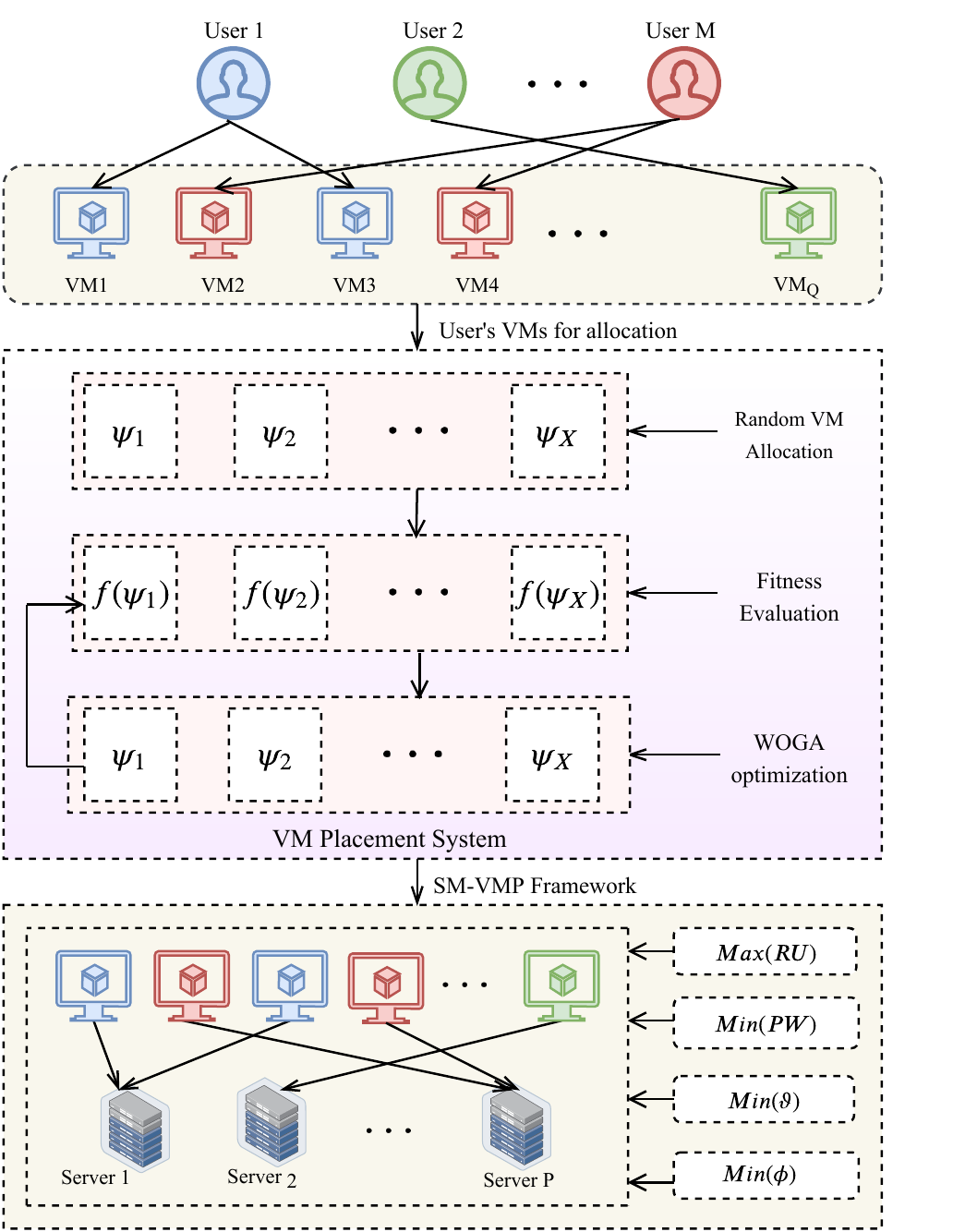}
	\caption{ SM-VMP Framework}
	\label{woga}
\end{figure}

 SM-VMP Framework consists of three consecutive steps namely, generation of random VM allocation, fitness evaluation and WOGA optimization as shown in Fig. \ref{woga}. Firstly, $X$ random feasible VM allocations (solutions) are generated. The $\psi_n$ represents $n^{th}$ possible VMP ($n \leq X$), where size of each $\psi$ vector is equal to the number of severs $P$. Each index value in $\psi$ vector represents placement of $j^{th}$ VM on randomly selected $i^{th}$ server satisfying all the constraints mentioned in Eqs. (\ref{1})-(\ref{4}) and updates server status to operational mode, $\omega_{ji}=1$. Secondly, {a cost function \{$\eta$\big($f_{RU_{dc}}(\psi_n)$, $f_{\phi_{dc}}(\psi_n)$, $f_{\vartheta_{dc}}(\psi_n)$, $f_{PW_{dc}}(\psi_n)$\big): $\forall_n\leq X$\} generating non-dominated pareto-front; is evaluated by computing four fitness values $RU_{dc}$, $\phi_{dc}$, $\vartheta_{dc}$, and $PW_{dc}$ associated with resource utilization (Eq. (\ref{ru})), security (Eq. (\ref{security})), communication cost (Eq. (\ref{comm})), and power consumption (Eq. (\ref{power2})), respectively. Thereafter, these solutions are sorted according to their dominance level and the non-dominated solutions are added into Pareto front. The $\psi_n$ dominates $\psi_m$ if the cost values of $\psi_n$ are better than $\psi_m$ on at least one objective and same or better on other objectives. The best solution among the multiple solutions in a pareto-front, is selected by computing crowding distance measured by adding objective-wise normalized difference between two neighbouring solutions of $n^{th}$ solution.} The solution having largest crowding distance is selected as best solution. In third step, WOGA optimization improves VMP using various operations including exploration and exploitation strategies of whale optimization, crossover and mutation operators of genetic algorithms followed by selection based on pareto-front. WOGA optimization is discussed in detail in the following subsection. %This optimization process is repeated till termination condition reaches. %Finally, VMs are allocated with feasible, optimal, secure and multi-objective virtual machine placement.

\subsection{WOGA Optimization}
\label{woga_optimization}
  VMs are encoded as whales and their allocations as whale positions, which are updated when VM migrates from one server to another by applying either encircling prey or random search strategy. \textit{Encircling prey} method shifts the current solution closer to the best solution by reducing distance between these two as stated in Eqs. (\ref{encircling prey1}) and (\ref{encircling prey2}), $\hat{D}$ is distance between best and current solutions, $\hat{A}$ and $\hat{C}$ are co-efficient vectors, $\psi_{best}^g$ is best solution available till $g^{th}$ iteration and $\psi^g$ and $\psi^{g+1}$ are current and updated solutions, respectively.
    \begin{equation}
  \hat{D} =|\hat{C} \times \psi_{best}^g - \psi^g| \label{encircling prey1}
  \end{equation}
  \begin{equation}
  \psi^{g+1} = \psi_{best}^g - \hat{A}\times\hat{D} \label{encircling prey2}
  \end{equation}
Likewise, \textit{random search for prey} is applied to explore the search space in multiple directions and update the VM allocations by allowing a random VM migration as stated in Eqs. (\ref{rand search for prey1}) and (\ref{rand search for prey2}): 
  \begin{equation}
  \hat{D}=|C \times \psi_{rand}^g - \psi^g| \label{rand search for prey1}
  \end{equation}
  \begin{equation}
  \psi^{g+1} = \psi_{rand}^g - \hat{A}\times\hat{D} \label{rand search for prey2}
  \end{equation}\\ 
  where, $\psi_{rand}^g$ and $\psi^g$ are random and current solutions respectively at $g^{th}$ iteration. The whale optimization updates VM allocations with improved fitness value which becomes parent chromosomes before application of GA operators. The detailed description of consecutive stages involved in WOGA optimization are as follow:

\begin{itemize}
\item In \textit{first stage}, the best whale position is determined by computing \textit{pareto-front} using non-dominated sorting based pareto-efficient optimization of NSGA-II. Thereafter, either WOA exploitation i.e., encircling prey (Eqs. (\ref{encircling prey1}) and \ref{encircling prey2}) or exploration i.e., random search for prey  operator (Eqs. (\ref{rand search for prey1}) and (\ref{rand search for prey2})) is applied based on the value of $\hat{A}$ decision variable by using Eqs. (\ref{shrink 1}) and (\ref{shrink 2}), 
\begin{equation}
\hat{A}= 2.\hat{x}.\hat{r} - \hat{x} \label{shrink 1}
\end{equation}
\begin{equation}
\hat{C}= 2.\hat{r} \label{shrink 2}
\end{equation}
 where, the variable $\hat{x}$ decreases linearly from 2 to 0 over the period of evolution, and {$\hat{r}$ is a random vector} in the range [0, 1]. The random search strategy raises diversity of the search space and prevents pre-mature convergence by applying Eqs. (\ref{rand search for prey1}) and (\ref{rand search for prey2}). The position of VMs are updated by migrating from non-optimal to near-optimal server depending upon the multi-objective fitness value.

\item In the \textit{second stage}, any infeasible solution having unallocated VMs, is turned into feasible solution by applying {First-Fit Decreasing (FFD)} strategy. {FFD \cite{jangiti2019aggregated} sorts the unallocated VMs in decreasing size, and apply the First-Fit VM packing algorithm to assign largest VM first to the available servers}. When feasible VM allocation is obtained, the search terminates and the process repeats. 

\item In the \textit{third stage}, updated whale position (VMP) vectors act as parent chromosomes with improved fitness values which undergoes crossover and mutation operations of GA. During crossover, two parent chromosomes $\psi_1$ and $\psi_2$ recombine at crossover-point to generate two new offspring $\psi_1^{child}$ and $\psi_2^{child}$. Therefore, population size gets double after crossover, which is followed by mutation operation which introduces sudden exchange of genes between two randomly selected positions within a chromosome and updates the VM allocation by migrating the VMs from non-optimal server to optimal server.
\item In \textit{fourth stage}, fitness value of each updated vector is evaluated and the fittest solution (or VM allocations) is selected as the best whale position. Finally, top $X$ solutions from pareto-optimal front are selected as improved feasible solutions. These four processes iterates till the termination condition reaches. The termination condition for the iterative improvement is either the number of iterations become greater than maximum iterations or when there is no improvement in successive iterations i.e., convergence point has reached. %The WOGA optimization based VMP is given in Algorithm \ref{euclid1}.
%\begin{figure}
\end{itemize}      
\subsubsection{Encoding of VM allocation as whale position vector}
 To encode VM allocation as whale position vectors, the probability values are used in the range [0, 1]. The best solution $\psi_{best}$ is represented as string of maximum probability value i.e., 1 as shown in first step of Fig. \ref{enco-decowoga}. The rest of the VM allocations are encoded with the values comparable to the probability values of best solution $\psi_{best}$ according to following criteria. For each solution $\psi_k$, we find out how much the $k^{th}$ VM allocation matches with the best VMP on the basis of the number ($V_{num}$) and types ($V_{t}$) of VMs on respective server to assign the probability values.
\begin{itemize}
	\item If both $V_{num}$ and $V_{t}$ of current solution exactly matches with that of best solution, then corresponding index is set to 1. 
	%\item When there is complete mismatch of type and number of VMs or there is no VM allocation on respective server, then the index value is 0.
	\item If $V_{t}$ is same and $V_{num}$ in current solution is half, one-third or one-fourth and so on, of the best vector, assign the probability values as 0.5, 0.33, 0.25,... respectively.
	\item If $V_{num}$ matches and $V_{t}$ is mismatched, then assign a value i.e., 1- degree of mismatch. For example, both best and current vectors have 4 VMs at $i^{th}$ server and 3 VM have different types, then probability value is $ 1-0.75=0.25$. 
	\item Otherwise, set index value to 0, since no VM is placed on the respective server.

\end{itemize}

 \subsubsection{Decoding of whale position vector as VM allocation} 
To decode probability values into actual VM allocation, the criteria is as follows:
\begin{itemize}
	\item If $\psi^{g+1}$ vector index value is 1, best vector VMs are copied (maintaining elitism) to the updated vector.
	\item If the value $\geq$ 0.75, then only one VM  from current is copied to the updated vector and rest allocations are copied from best vector.
	
	\item  If the value is between 0.50 and 0.75, then first 50\% of VMs are allocated from best vector and remaining 50\% are copied from last VM allocation of current vector.
	\item If the value $>$ 0.25 and $\leq$ 0.50, then 75\% VMs are copied from current vector and 25\% from best vector.
	\item  Otherwise, all VMs are copied from current vector to the updated vector.
	
\end{itemize} 
\begin{figure}[!htbp]
	\centering
	\includegraphics[width=0.99\linewidth]{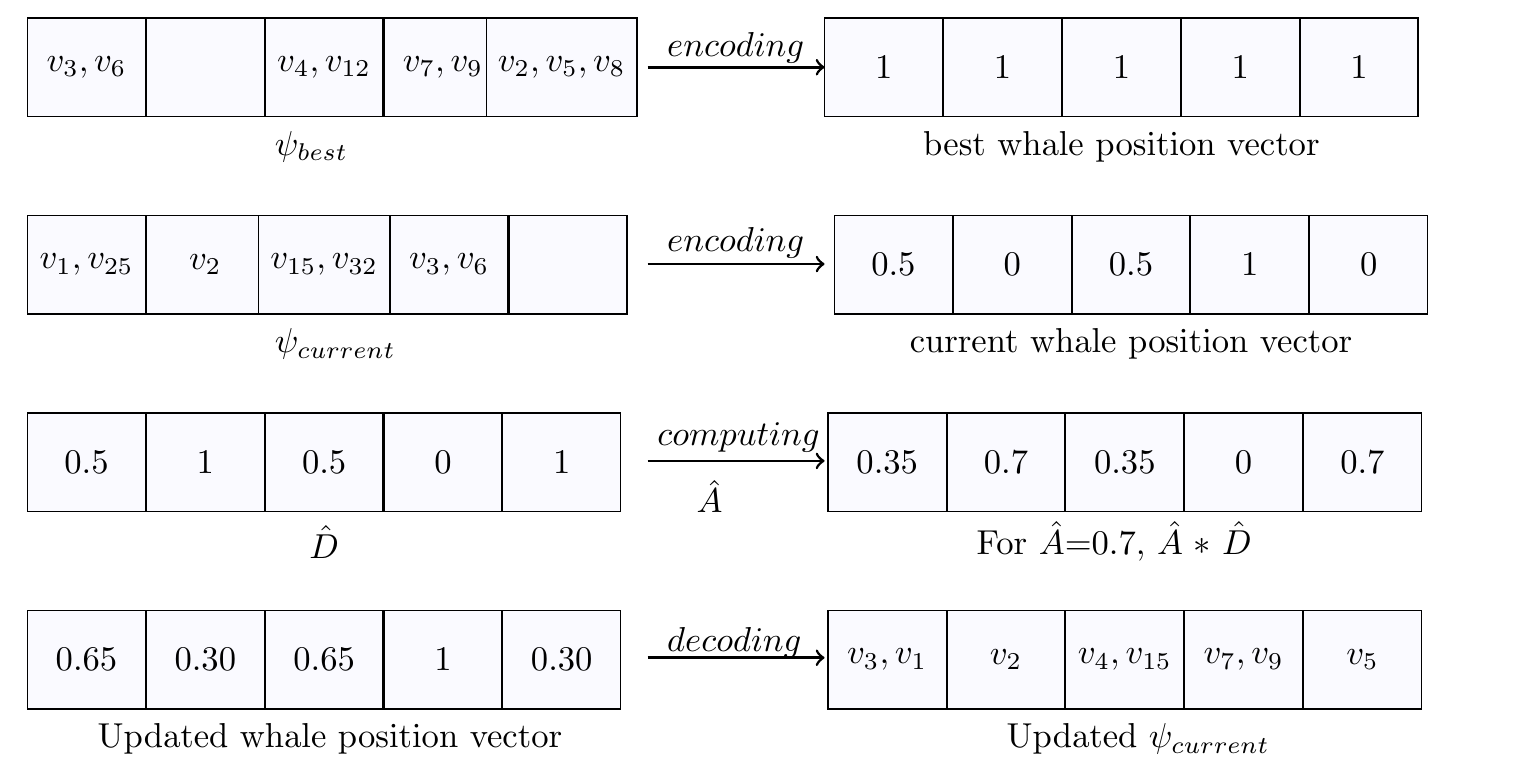}
	\caption{Encoding and decoding for whale optimization}
	\label{enco-decowoga}
\end{figure}

 \textit{For instance}, there are four different types of VMs depending on variation of CPU, storage and RAM capacity. The VMs $v_1$ to $v_{10}$, $v_{11}$ to $v_{20}$, $v_{21}$ to $v_{30}$ and $v_{31}$ to $v_{50}$ belong to Type-I, Type-II, Type-III, and Type-IV VMs, respectively. The best and current whale positions are shown in first and second steps of Fig. \ref{enco-decowoga}, where server $1$ hosts two VMs $v_3,v_{6}$ in best and $v_1,v_{25}$ in current vector, where number and 50\% type of VMs matches as $v_3,v_{6},v_1$ belongs to Type-I VM and $v_{25}$ is of different type, hence, the value $0.5$ is assigned to index 1 of current vector and so on. Thereafter, difference between positions of best and current whale vectors, denoted as $\hat{D},$ is computed as shown in third step of Fig. \ref{enco-decowoga}. The optimization achieved by applying encircling prey strategy (given in Eqs. (\ref{encircling prey1}) and (\ref{encircling prey2})), is shown in the fourth step of Fig. \ref{enco-decowoga}, where updated whale position $\psi_{n+1}$ is decoded into next VM allocation vector. In case of random strategy, the best vector is replaced by the random vector while following the above mentioned encoding and decoding steps.
 
\subsection{Operational Design and Complexity Computation}
Algorithm \ref{euclid1} describes the operational summary of SM-VMP. Step 1 initializes list of VMs ($List_{VM}$), list of servers ($List_{PM}$), maximum number of iterations ($Gmax$), and iteration counter ($g$). The steps 2-16 generate $X$  VMP allocations (or solutions) randomly whose time complexity depends upon $X$, $P$ and $Q$ equals to $O(XPQ)$. The steps 6-14 specify that each allocation must satisfy the resource capacity constraints mentioned in Eqs. (\ref{1})-(\ref{4}). The steps 17-19 remove unallocated VMs from $k^{th}$ solution ($\psi_k$) and refill them by applying FFD algorithm having 
\begin{figure}[!htbp]
	\removelatexerror  
	\begin{algorithm}[H]
		\caption{ SM-VMP Operational Summary}
		\label{euclid1}
		%\begin{algorithmic}[1]
		Initialize: $g \leftarrow 0$, $Gmax$, $X$, $List_{VM}$, $List_{PM}$\; 
		\For {each $k = (1,2,...,X)$} {
			%\For {$i={1,2, ..., Q}$}{
			\While {$List_{VM} \neq Empty$} 
			{Choose $v_j$ from $List_{VM}$ randomly\;
				{Choose $S_i$ from $List_{PM}$ randomly}\;
				\For {each $\mathds{R} \in \{CPU, RAM, storage, PE\} $} {
					
					\eIf {$v_j^{\mathds{R}} \leq S_i^{\mathds{R}} $} {
						Assign $v_j$ to $S_i$ and update $\omega_{ji}=1$ and  mapping vector $\psi_{k}$\;
						Update $S_i^{\mathds{R}}= S_i^{\mathds{R}} - v_j^{\mathds{R}} $\;
						Discard $v_j$ from $List_{VM}$\;
					
					}
					{ $v_j$ is not allocated and $\omega_{ji}=0$\;}
				}
			}
		
		}
		\For {$k={1,2,..., X}$}{
			Call First-Fit Decreasing (FFD) algorithm for the unallocated VMs and update mapping vector $\psi_{k}$\;}
		%\For {$j={1,2,..., Q}$}{			
		%\eIf {$v_j$ is unallocated}{    
		%	[$\psi_{k}$, $\omega_{ji}$]= First-Fit Decreasing($v_j$) \;	} 
		%{Keep ${v}_j$ at same server until user terminates it \;
		%} 
	
		\While { termination criteria}{
			
			%\STATE Declare $C = []$
			%\FOR {each i=(1,2,...,X) }
		[$f_1(RU_g)$, $f_2(\phi_g)$, $f_3(\vartheta_g)$, $f_4(PW_g)$] = $\eta(\psi^g)$\;
			$\psi_{best} \leftarrow NDS$($f_{RU_{dc}}(\psi_n)$, $f_{\phi_{dc}}(\psi_n)$, $f_{\vartheta_{dc}}(\psi_n)$, $f_{PW_{dc}}(\psi_n)$)\; %$\psi_{best} \leftarrow NDS(\psi^g)[0]$\;
			%$[\psi_{sorted}= Pareto\_front(\psi^g)]$, $\psi_{best} \leftarrow \psi_{sorted}[0]$\;
		
			$\psi^g_{updated}$ = Whale Optimization($\psi^{g}$, $\psi_{best}$) \;
			\For {$k={1,2,..., X}$}{
				Call FFD algorithm for the unallocated VMs and update mapping vector $\psi^g_{updated}$\;}
			\For {each k=(1,2,...,X) }{
				$rn= random(1,X)$, $cp= random(1,P)$\;
				
				$C_1 = [\psi^g_k(1:cp), \psi^g_{rn}(cp +1:P)]$\;
				$C_2 = [\psi^g_{rn}(1:cp), \psi^g_{k}(cp +1:P)]$\;
				$C= [C, \mu(C_1), \mu(C_2)]$\;
				
			Call FFD algorithm for the unallocated VMs and update vector $C$\;  	
			%	$VM^{feasible}$=First-Fit Decreasing($C$) \;
			
			}	
			 [$f_1(RU_g)$, $f_2(\phi_g)$, $f_3(\vartheta_g)$, $f_4(PW_g)$ ]= $\eta(C)$\;
			$\psi^g=[\psi^g,C]$; $\psi^{g+1}$= $NDS$($f_{RU_{dc}}(\psi^g)$, $f_{\phi_{dc}}(\psi^g)$, $f_{\vartheta_{dc}}(\psi^g)$, $f_{PW_{dc}}(\psi^g)$)\;		
		}	
		%\end{algorithmic}
	\end{algorithm} 
\end{figure}
time complexity: $O(XPQ)$. Steps 20-36 repeat till accomplishment of termination criteria (by utilizing \textit{fourth stage} of Section \ref{SM-VMP}).\ref{woga_optimization}), where step 21 evaluates cost values of four objectives using Eqs. (\ref{ru}), (\ref{security}), (\ref{comm}), and (\ref{power2}) associated to each solution of $g^{th}$ iteration (i.e., $\psi^{g}$). Step 22 selects the best solution ($\psi_{best}$) from the pareto-front generated by applying Non-Dominated Sorting ($NDS$) \cite{deb2002fast} with time complexity of $O((ob)X^2)$, where $ob$ is number of objectives. Similarly, step 23 calls whale optimization operators having complexity of $O(PX)$ (by following \textit{first stage} of Section \ref{SM-VMP}.A). Steps 24-26 call FFD to assign unallocated VMs (by utilizing \textit{second stage} of Section \ref{SM-VMP}.\ref{woga_optimization}). {Steps 27-33 explore the solution space for a better allocation by recombining the individuals using single-point crossover which generates less infeasible solutions and mutation operation that helps in exploring the entire search space while avoiding local optima (by following Section \ref{SM-VMP}.\ref{woga_optimization}, \textit{third stage}). The crossover point ($cp$) is randomly generated in step 28 and two offsprings ($C_1$, $C_2$) are generated in steps 29 and 30. Further, mutation operation ($\mu$) is applied to offsprings for production of updated solution vector ($C$) in step 31}. Again, all the unallocated VMs are removed and FFD is called to produce feasible solutions in step 32. The updated solutions are produced by evaluating cost function ($\eta(C)$) in step 34 and generating pareto-front by applying $NDS$ in step 35. The crossover and mutation  require $O(XP^2)$ and $O(XP)$ comparisons respectively. Therefore, the total time complexity is $O((ob)X^2P^2Q(Gmax))$.

\subsection{Illustration} Consider a data centre having four servers of two different types. Two servers belongs to Type-I having configuration CPU of 1000 MIPS, 1200 GB of memory and maximum power consumption ${PW_i}^{max}$ =844 W, ${PW_i}^{min}$=${PW_i}^{idle}$= 120 W  and remaining two servers are of Type-II with CPU of 1500 MIPS, 1500 GB memory and ${PW_i}^{max}$ =1024 W, ${PW_i}^{min}$=${PW_i}^{idle}$= 160 W  power consumption. Let three users requested eight VMs which are deployed on these four servers. The CPU and memory utilization of eight VMs are as follows: (200, 250), (250, 310), (400, 350), (150, 200), (600, 650), (180, 200), (450, 500), (100, 150). There are $8^4$ feasible or infeasible ways to place these VMs on four servers. At an instance, random VM allocation is as follows: $v_{11}^1$, $v_{22}^1$,$v_{34}^1$, $v_{43}^1$, $v_{51}^2$ , $v_{63}^2$, $v_{74}^2$, $v_{83}^3$ where $v_{ij}^k$ denotes $i^{th}$ VM is deployed on $j^{th}$ server owned by $k^{th}$ user. The resource utilization $RU_{dc}$, power consumption, conflicting servers, and communication cost are 47.2\%, 1.49E+05 W, 75\%, and 66.67\% respectively. The application of WOGA algorithm optimizes the above mentioned VM allocation as  $v_{13}^1$, $v_{23}^1$,$v_{33}^1$, $v_{43}^1$, $v_{54}^2$, $v_{63}^2$, $v_{74}^2$, $v_{82}^3$ by keeping server $1$ in sleep mode with resource utilization, power consumption, conflicting servers, and communication cost up to 58.2\%, 1.42E+05 W, 25\%, and 33.6\%  respectively. It is worth noting that in proposed VM allocation, the conflicting servers and communication cost are reduced by 66.6\% and 50\% respectively, with an improved resource utilization and decreased power consumption.

\section{Multi-efficient VM migration}
The VMs are re-allocated subsequently on arrival of new request and de-allocated on completion of respective tasks, to handle dynamic workload demands of cloud users. This gives an opportunity to re-optimize the VMP dynamically, by migrating the VMs from under/over-utilized servers to selected efficient server machines. Since, we are placing VMs with cloud user as well as service provider's perspectives, the migrating VM is checked for its type: compute-sensitive or data-sensitive through its configuration as if the required CPU capacity $\ge$ memory capacity, then it is compute-sensitive, otherwise, data-sensitive. For compute-sensitive VM, migration to some active server is tried first, instead of switching on an inactive server and the server having largest resource capacity is activated only if the migrating VM creates overload on already active servers. The purpose behind activating largest capacity server is to minimize the number of active servers and delaying need to turn-on another inactive server to accommodate future VM migrations. On the other hand, for migrating highly data-intensive VM, switching to near-by located active server is tried first. For experimental purpose, we assume that VMs of common user are inter-dependent which are intentionally allocated on servers that are placed closer physically. However, the decision of compute or communication sensitive VMs can be taken by user or data analyst at data centre.
% Hence, if the migrating VM belongs to same user, then keep the respective VM on same server, and shift some other user's VM from over-loaded server so as to allow communication and energy-efficient VM migration. 
 To accomplish the same, the VMs are migrated from the server (which become under-utilized on the release of VMs) to near-by optimal server and idle servers are shutdown. %This approach of VM migration allows consolidation of VMs with reduced power consumption or network communication cost according to their type.
  The total migration cost $MG_{cost}$ is evaluated using Eq. (\ref{migration});
\begin{equation}\label{migration}
\resizebox{0.4\textwidth}{!}{$  
	\begin{aligned}
	MG_{cost}= \sum{h_{ij}*(\mathds{D}(S_k ,S_j)*\mathds{H}(v_i))} + \sum{n_{j}*{z_{j}}}
	%\quad \forall i \in O_v, j \in S 
	\end{aligned}$}
\end{equation}
where $ \forall i \in O_v, j \in P$, $ \mathds{D}(S_k ,S_j)$ is the distance or number of hops covered by migrating VM ($v_i$) from source $S_k$ to destination server $S_j$, $\mathds{H}(v_i)$=$v_{cpu} \times v_{memory}$, is the size of $i^{th}$ migrating VM, $O_v$ is the list of VMs on overloaded server ($S_k$). The first term $\sum{h_{ij}*{\mathds{D}(S_k ,S_j) *\mathds{H}(v_i)}} $ signifies networking cost consumed for migration. Since, in proposed approach, VMs of common user are preferably placed closer, the server located at minimum distance from overloaded server is selected as destination server to optimize VM migration cost. The second term $\sum{n_{j}*{z_{j}}}$ specifies server state transition energy where, if $i^{th}$ VM is placed at $j^{th}$ server after migration, then $h_{ij}=1$, otherwise, it is 0. If $j^{th}$ server receives one or more VMs after migration, then $n_{j}=1$ else it is 0. Similarly, if $z_{j}=0$ means $j^{th}$ server is already active before migration, otherwise, $z_{j}=E$ where $E$ is energy consumed in switching a server from idle to active state, which is equals to $4260$ Joules as stated in \cite{dabbagh2015exploiting}. Eq. (\ref{c1}) states that each VM must be assigned to only one server. Eq. (\ref{c2}) specifies that resource requirement of migrating VM must be lesser than respective available resource capacity of the selected destination server. 
\begin{gather}
	\sum_{j \in P}{c_{ij}} =1 \quad \forall_i \in O_{v} \label{c1}
\end{gather}

\begin{equation}
	\sum_{i \in O_v}{c_{ij}v_i^{\mathds{R}}} \leq S_j^{\mathds{R}} \quad \forall_j \in P, \mathds{R} \in \{C, Mem, RAM\}
	\label{c2}
\end{equation}

\section{Performance Evaluation }

\subsection{Experimental Set-up}
The simulation experiments are executed on a server machine assembled with two Intel\textsuperscript{\textregistered} Xeon\textsuperscript{\textregistered} Silver 4114 CPU with 40 core processor and 2.20 GHz clock speed. The computation machine is deployed with 64-bit Ubuntu 16.04 LTS, having main memory of 128 GB. The data centre environment consisting of a cluster of servers arranged in a linear topology was set up in Python with three different types of servers and four types of VMs configuration shown in Tables \ref{table:server} and \ref{table:vm}, respectively. The features like power consumption ($P_{max}, P_{min}$), MIPS, RAM and memory are taken from real server configuration: IBM \cite{IBM1999} and Dell \cite{Dell1999}, where $S_1$, $S_2$ and $S_3$ are `ProLiantM110G5XEON3075', `IBMX3250Xeonx3480' and `IBM3550Xeonx5675', respectively. The VMs configurations are inspired from the VM instances of Amazon website \cite{amazon1999EC2}. The dynamically generated users execute applications in the form of Bag of Tasks and the number of tasks may vary in the range [1, 10] generated randomly at run-time. Each user holds VMs depending upon the number of tasks with a constraint that total number of VM requests must not exceed the size of data centre and each VM belongs to atmost one user. Therefore, each user's application is shared among random number of multiple VMs.
 
\begin{table}[htbp]
	\centering
	
	\caption[Table caption text] {Server configuration}  %\cite[p.10]{refid} }
	\label{table:server}
		\resizebox{0.9\textwidth}{!}{\begin{minipage}{\textwidth}
	\begin{tabular}{lcccccc}
		\hline
		%\multicolumn{2}{c}{Item} \\
		%\cline{1-2}
		Server&PE&MIPS& RAM(GB) & Storage(GB) &$P_{max}$ &$P_{min}$/$P_{idle} $\\
		\hline
		$S_1$& 2&2660&4&160&135&93.7 \\
		$S_2$& 4&3067&8&250&113&42.3 \\
		$S_3$& 12&3067&16&500&222&58.4 \\
		
		%Mutation learning period( $lp^M$)& 10\\
		%Crossover rate learning period($lp^{CR}$) & 10\\
		
		\hline
	\end{tabular}
		\end{minipage}}
\end{table}

\begin{table}[htbp]
	\centering
	
	\caption[Table caption text] {VM configuration}  %\cite[p.10]{refid} }
	\label{table:vm}
	%	\resizebox{0.8\textwidth}{!}{\begin{minipage}{\textwidth}
	\begin{tabular}{lcccc}
		\hline
		%\multicolumn{2}{c}{Item} \\
		%\cline{1-2}
		VM type& PE &MIPS&RAM(GB)&Secondary storage(GB)\\
		\hline
		$VM_{S}$&1&500&0.5&40\\
		$VM_{M}$&2&1000&1&60\\
		$VM_{L}$&3&1500&2&80\\
		$VM_{XL}$&4&2000&3&100\\

		\hline
	\end{tabular}
	%	\end{minipage}}
\end{table}
 
\subsection{Simulation Results}
Table \ref{table:woga} analyses the performance of proposed work, over different data centre simulations, which includes various combination of VMs and servers. Each experiment is executed for 25 times and mean of the obtained results are reported. Since the number of users are dynamic in real cloud environment, therefore, to analyse the percentage of conflicting server and communication cost, the experiments are conducted with 10, 20, ..., 60 users for 100 VMs as shown in Table \ref{table:users_100VMs}. With various number of users different outputs of conflicting server and communication cost were attained. It is observed that with increase in the number of users, the percentage of conflicting servers increase and communication cost slightly decreases, because total number of VMs are fixed and user could own limited number of VMs. However, the most promising outcomes are obtained with 40 users for 100 VMs where conflicting server as well as communication cost percentage produce non-dominant results. Moreover, the mean absolute deviation of the obtained performance metrics which are independent of size of data centre including $RU$, $\phi$, and $\vartheta$ are shown in Table \ref{table:woga} and the achieved standard deviation of resource utilization, conflicting servers (\%) and communication cost (\%) are 0.024, 1.49 and 0.47 respectively, which are acceptable as compared to their obtained respective values. From the standard deviation, it can be analysed that WOGA allows stable performance. 
\begin{table}[!htbp]
	\centering
	
	\caption[Table caption text] {Performance metrics for SM-VMP}  %\cite[p.10]{refid} }
	\label{table:woga}
	\resizebox{0.88\textwidth}{!}{\begin{minipage}{\textwidth}
			\begin{tabular}{lcccccccc}
				\hline
				%\multicolumn{2}{c}{Item} \\
				%\cline{1-2}
				VMs(PMs)&User & $RU$(\%)&$\sigma_{RU}$&$PW$ ($W$)&$\phi$(\%)&$\sigma_{\phi}$&$\vartheta$(\%)&$\sigma_{\vartheta}$\\
				\hline
				50(30)& 20&63.98&0.46  &3.80E+3 &15.78 & 0.60   &12.34& 4.60  \\
				100(60)&40& 63.57& 0.05  &7.69E+3 &14.34& 2.07   & 11.42& 5.54 \\
				200(120)&80&63.05&0.47  &1.54E+4 &13.23& 3.15   &16.92& 0.02  \\
				%300(180)&62.34&2.31E+04&15.34&18.28&852.86\\
				400(240)&160&62.87&0.65 &3.09E+4 &17.32& 0.94  &17.14& 0.20 \\
				%500(300)&60.32&3.88E+04&13.25&18.45&1465.61\\
				600(360)&240&63.19&0.33  &4.63E+4  &16.42& 0.02   &20.10&  3.16 \\
				800(480)&320&64.07& 0.55  &5.17E+4  &18.51&   2.12  &20.81&  3.87 \\
				1000(600)&400&63.95&0.43  &5.95E+4   &19.10&  2.72  &19.89& 2.95 \\
				
				\hline
			\end{tabular}
	\end{minipage}}
\end{table}
\begin{table}[!htbp]
	\centering
	\caption[Table caption text] {Performance metrics for 100 VMs and 60 servers}   %\cite[p.10]{refid} }
	\label{table:users_100VMs}
	%\resizebox{0.8\textwidth}{!}{\begin{minipage}{\textwidth}
	\resizebox{5.5cm}{!}{
		\begin{tabular}{lcccc}
			\hline			
			{User \#} &{$RU$(\%)}&{$PW$ ($W$)} & {$\phi$(\%)}&{$\vartheta$(\%)} \\ 	\hline
		{10} &{63.45} &{7.70E+3} & {1.22}& {30.00}\\
			{20} &{63.23} &{7.63E+3} &{5.42} &{19.87} \\
			{30} &{62.97} &{7.65E+3} &{8.86} & {13.34}\\
			{40} &{63.57} &{7.69E+3} &{14.34} & {11.42}\\
			{50} &{63.25} &{7.69E+3} & {20.19}& {12.67}\\
			{60} &{63.03} &{7.68E+3} &{23.60} &{11.66} \\
			
			\hline	
	\end{tabular}}
	%\end{minipage}}
\end{table}
 It is observed in Table \ref{table:woga} that the resource utilization is closer to 64\% for each data centre while power consumption and number of active servers changes with respect to the number of VMs. The number of conflicting servers ($\phi$) and the communication cost ($\vartheta$) are obtained in the range of {(13\%-20\%)} and (11\%-21\%), respectively. The time of execution is {linear} with respect to the number of requested VMs (shown in Fig. \ref{staticVMP}(f)).

\subsection{Comparative Analysis}
The state-of-the-arts selected for the comparison are \textit{HGAPSO} \cite{sharma2016multi}, \textit{NSGA-II} \cite{guerrero2018multi}, \textit{GA} \cite{tseng2017dynamic}, \textit{WOA} \cite{rana2018cloud}, \textit{SMOOP} \cite{han2017reducing}, \textit{First-Fit} \cite{jangiti2019aggregated}, \textit{Best-Fit} \cite{shirvastava2017best} and \textit{Random-Fit} \cite{jung2010mistral}. These works and their relative comparison are already discussed in related work. Though VMP is presented with two or three objectives only in these works, we have implemented their approaches and compared thoroughly for all the four objectives (by utilizing our proposed objective models wherever required) in two different scenarios viz. Static VMP and Dynamic VMP.
\subsubsection{ Static VMP}
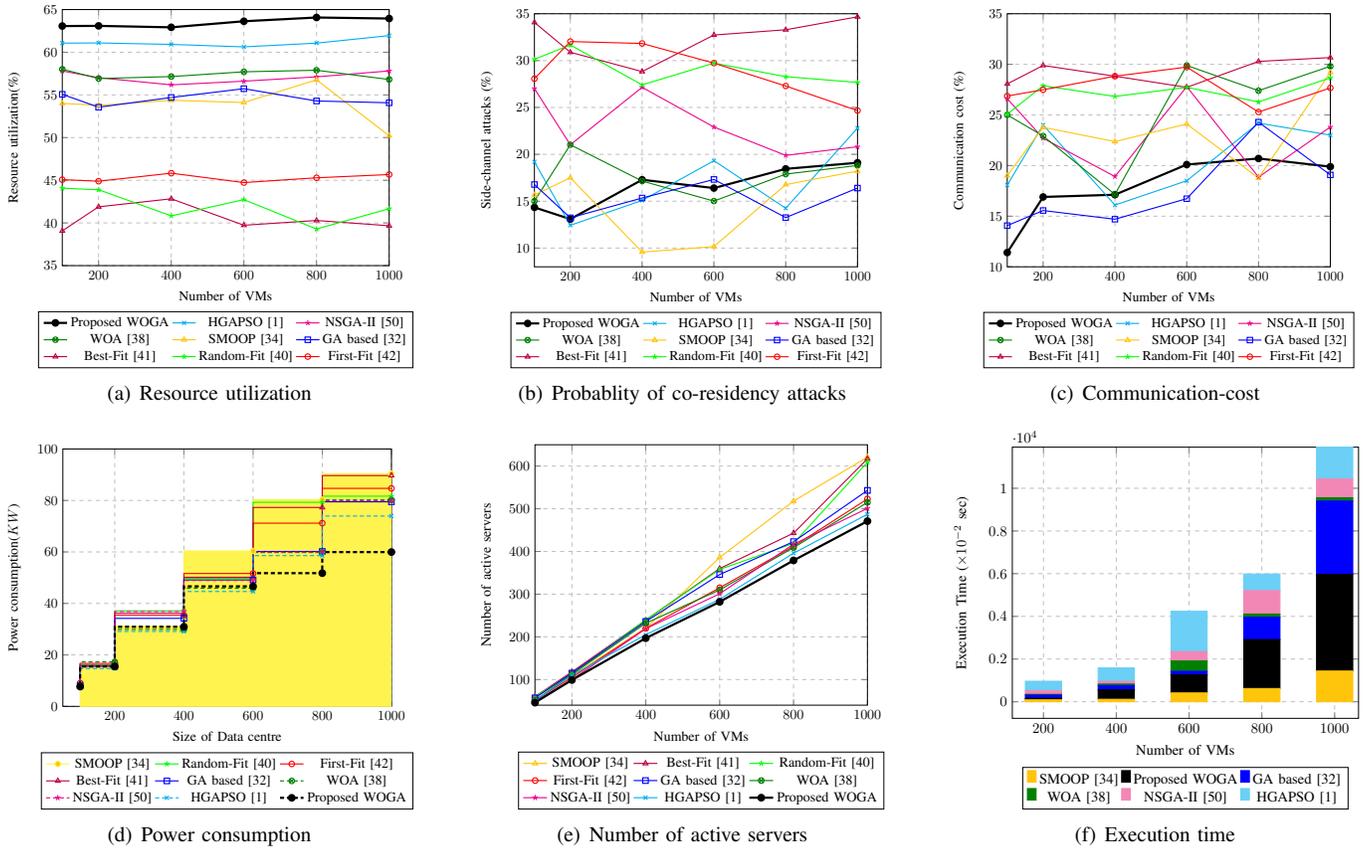
\begin{figure*}[htbp!]
	
	\centering
	\subfigure[Resource utilization ]{%\includegraphics[width=.315\textwidth]{Results/ru}
		\resizebox{0.3\textwidth}{!}{\begin{tikzpicture}
		\begin{axis}[
		width=0.55\textwidth,
		height=0.45\textwidth,
		%axis lines=left,
		%axis on top=false,
		xmin=100,
		xmax=1000,
		xlabel={Number of VMs},
		xticklabel style={/pgf/number format/1000 sep=},
		ymin=35,
		ymax=65,
		ylabel={Resource utilization(\%)},
		legend style={at={(0.5,-0.4)},
			anchor=south ,legend columns=3},
		ymajorgrids=true,
		xmajorgrids=true,
		grid style=dashed,
		]
		\addplot[color=black,mark=oplus,ultra thick] coordinates {
			(1000,63.95) (800,64.07) (600,63.63) (400,62.92) (200,63.09) (100,63.07)
		};\addlegendentry{Proposed WOGA}
		\addplot[color=cyan,mark=x,thick] coordinates {
			(1000,61.95) (800,61.07) (600,60.63) (400,60.92) (200,61.09) (100,61.07)
		};\addlegendentry{HGAPSO \cite{sharma2016multi}}
		
		\addplot[color=magenta,mark=star,thick] coordinates {
			(1000,57.81) (800,57.12) (600,56.61) (400,56.18) (200,57.02) (100,57.81)
		};\addlegendentry{NSGA-II \cite{guerrero2018multi}}
		\addplot[color=mypink,mark=otimes,thick] coordinates {
			(1000,56.81) (800,57.89) (600,57.71) (400,57.15) (200,56.92) (100,58.01)
		};\addlegendentry{WOA \cite{rana2018cloud}}
		\addplot[color=Mycolor,mark=triangle,thick] coordinates {
			(1000,50.21) (800,56.77) (600,54.11) (400,54.37) (200,53.78) (100,54.01)
		};\addlegendentry{SMOOP \cite{han2017reducing}}
		\addplot [color=blue,mark=square,thick] coordinates { 
			(1000,54.08) (800,54.29) (600,55.73) (400,54.7) (200,53.56) (100,55.07)};\addlegendentry{GA based \cite{tseng2017dynamic}}
		\addplot [color=purple,mark=triangle,thick] coordinates { 
			(1000,39.67) (800,40.29) (600,39.73) (400,42.83) (200,41.89) (100,39.07)};\addlegendentry{Best-Fit \cite{shirvastava2017best}}
		\addplot [color=green,mark=star,thick] coordinates { 
			(1000,41.67) (800,39.29) (600,42.73) (400,40.83) (200,43.89) (100,44.07)};\addlegendentry{Random-Fit \cite{jung2010mistral}}
		\addplot [color=red,mark=o,thick] coordinates { 
			(1000,45.67) (800,45.29) (600,44.73) (400,45.83) (200,44.89) (100,45.07)};\addlegendentry{First-Fit \cite{jangiti2019aggregated}}
		
		\end{axis}
		\end{tikzpicture}}
		
	}\hfill
	\subfigure[Probablity of co-residency attacks ]{%\includegraphics[width=.315\textwidth]{Results/sca}
	\resizebox{0.3\textwidth}{!}{\begin{tikzpicture}
		\begin{axis}[
		width=0.55\textwidth,
		height=0.45\textwidth,
		%axis lines=left,
		%axis on top=true,
		xmin=100,
		xmax=1000,
		xlabel={Number of VMs},
		xticklabel style={/pgf/number format/1000 sep=},
		ymin=8,
		ymax=35,
		ylabel={Side-channel attacks (\%)},
		legend style={at={(0.5,-0.4)},
			anchor=south ,legend columns=3},
		ymajorgrids=true,
		xmajorgrids=true,
		grid style=dashed,
		]
		\addplot[color=black,mark=oplus,ultra thick] coordinates {
			(1000,19.10) (800,18.45) (600,16.4) (400,17.286) (200,13.09) (100,14.34)  
			
		};\addlegendentry{Proposed WOGA}
		\addplot[color=cyan,mark=x,thick] coordinates {
			(1000,22.8) (800,14.25) (600,19.32) (400,15.08) (200,12.43) (100,19.2)
		};\addlegendentry{HGAPSO \cite{sharma2016multi}}
		
		\addplot[color=magenta,mark=star,thick] coordinates {
			(1000,20.81) (800,19.89) (600,22.91) (400,27.15) (200,21.02) (100,27.01)
			
		};\addlegendentry{NSGA-II \cite{guerrero2018multi}}
		\addplot[color=mypink,mark=otimes,thick] coordinates {
			(1000,18.81) (800,17.89) (600,15.01) (400,17.15) (200,21.02) (100,15.01)
		};\addlegendentry{WOA \cite{rana2018cloud}}
		\addplot[color=Mycolor,mark=triangle,thick] coordinates {
			(1000,18.21) (800,16.77) (600,10.16) (400,9.57) (200,17.5) (100,15.66)
		};\addlegendentry{SMOOP \cite{han2017reducing}}
		\addplot [color=blue,mark=square,thick] coordinates { 
			(1000,16.4) (800,13.25) (600,17.32) (400,15.34) (200,13.23) (100,16.78) 
		};\addlegendentry{GA based \cite{tseng2017dynamic}}
		\addplot [color=purple,mark=triangle,thick] coordinates { 
			(1000,34.67) (800,33.29) (600,32.73) (400,28.83) (200,30.89) (100,34.07)};\addlegendentry{Best-Fit \cite{shirvastava2017best}}
		\addplot [color=green,mark=star,thick] coordinates { 
			(1000,27.67) (800,28.29) (600,29.73) (400,27.41) (200,31.67) (100,30.14)};\addlegendentry{Random-Fit \cite{jung2010mistral}}
		\addplot [color=red,mark=o,thick] coordinates { 
			(1000,24.67) (800,27.29) (600,29.73) (400,31.83) (200,32.03) (100,28.07)};\addlegendentry{First-Fit  \cite{jangiti2019aggregated}}
		
		\end{axis}
		\end{tikzpicture}

}
         }\hfill
	\subfigure[Communication-cost ]{%\includegraphics[width=.315\textwidth]{Results/Com_cost}
	\resizebox{0.3\textwidth}{!}{\begin{tikzpicture}
		\begin{axis}[
		width=0.55\textwidth,
		height=0.45\textwidth,
		%axis lines=left,
		%axis on top=true,
		xmin=100,
		xmax=1000,
		xlabel={Number of VMs},
		xticklabel style={/pgf/number format/1000 sep=},
		ymin=10,
		ymax=35,
		ylabel={Communication cost (\%)},
		legend style={at={(0.5,-0.4)},
			anchor=south ,legend columns=3},
		xmajorgrids=true,
		ymajorgrids=true,
		grid style=dashed,
		]
		\addplot[color=black,mark=oplus,ultra thick] coordinates {
			(1000,19.89) (800,20.7) (600,20.10) (400,17.12) (200,16.9) (100,11.407)
		};\addlegendentry{Proposed WOGA}
		\addplot[color=cyan,mark=x,thick] coordinates {
			(1000,23.01) (800,24.2) (600,18.5) (400,16.1) (200,24.02) (100,18.07)
		};\addlegendentry{HGAPSO \cite{sharma2016multi}}
		
		\addplot[color=magenta,mark=star,thick] coordinates {
			(1000,23.81) (800,18.86) (600,27.8) (400,18.92) (200,22.7) (100,26.6)
		};\addlegendentry{NSGA-II \cite{guerrero2018multi}}
		\addplot[color=mypink,mark=otimes,thick] coordinates {
			(1000,29.81) (800,27.4) (600,29.89) (400,17.15) (200,22.92) (100,25.01)
		};\addlegendentry{WOA \cite{rana2018cloud}}
		\addplot[color=Mycolor,mark=triangle,thick] coordinates {
			(1000,29.21) (800,18.77) (600,24.11) (400,22.37) (200,23.78) (100,19.01)
		};\addlegendentry{SMOOP \cite{han2017reducing}}
		\addplot [color=blue,mark=square,thick] coordinates { 
			(1000,19.08) (800,24.29) (600,16.73) (400,14.7) (200,15.56) (100,14.07)};\addlegendentry{GA based \cite{tseng2017dynamic}}
		\addplot [color=purple,mark=triangle,thick] coordinates { 
			(1000,30.67) (800,30.29) (600,27.73) (400,28.83) (200,29.89) (100,28.07)};\addlegendentry{Best-Fit \cite{shirvastava2017best}}
		\addplot [color=green,mark=star,thick] coordinates { 
			(1000,28.67) (800,26.29) (600,27.73) (400,26.83) (200,27.89) (100,25.07)};\addlegendentry{Random-Fit \cite{jung2010mistral}}
		\addplot [color=red,mark=o,thick] coordinates { 
			(1000,27.67) (800,25.29) (600,29.73) (400,28.83) (200,27.49) (100,26.87)};\addlegendentry{First-Fit \cite{jangiti2019aggregated}}
		
		\end{axis}
		\end{tikzpicture}
	}
}
	\subfigure[Power consumption ]{%\includegraphics[width=.315\textwidth]{Results/power}
	\resizebox{0.3\textwidth}{!}{\begin{tikzpicture}
		\begin{axis}[
		width=0.55\textwidth,
		height=0.45\textwidth,
		%axis lines=left,
	%	axis on top=true,
		xmin=50,
		xmax=1000,
		xlabel={Size of Data centre},
		xticklabel style={/pgf/number format/1000 sep=},
		ymin=0,
		ymax=100,
		ylabel={Power consumption($KW$)},
		legend style={at={(0.5,-0.4)},
			anchor=south ,legend columns=3},
		xmajorgrids=true,
		ymajorgrids=true,
		grid style=dashed,
				]
		\addplot+[const plot, fill=Mycolor][color=yellow!80!, mark options={fill=Mycolor} ,thick] coordinates {
			(1000,90.41) (800,80.27) (600,60.37) (400,31.05) (200,15.68) (100,7.77)  }\closedcycle;\addlegendentry{SMOOP \cite{han2017reducing}}
		\addplot+[const plot] [color=green,mark=star,thick] coordinates { 
			(1000,81.67) (800,79.29) (600,49.73) (400,37.1) (200,16.1) (100,7.97)};\addlegendentry{Random-Fit \cite{jung2010mistral}}
		\addplot+[const plot] [color=red,mark=o,thick] coordinates { 
			(1000,84.7) (800,71.19) (600,51.63) (400,35.38) (200,16.69) (100,9.07)};\addlegendentry{First-Fit \cite{jangiti2019aggregated}}
		\addplot+[const plot] [color=purple,mark=triangle,thick] coordinates { 
			(1000,89.67) (800,77.29) (600,50.173) (400,36.183) (200,15.89) (100,7.78)};\addlegendentry{Best-Fit \cite{shirvastava2017best}}
		
		\addplot+[const plot] [color=blue,mark=square,thick] coordinates { 
			(1000,79.48) (800,60.19) (600,49.13) (400,34.2) (200,15.56) (100,8.07)};\addlegendentry{GA based \cite{tseng2017dynamic}}
		\addplot+[const plot][color=mypink,mark=otimes,thick] coordinates {(1000,80.21) (800,59.77) (600,45.91) (400,29.9) (200,17.28) (100,7.99)
			
		};\addlegendentry{WOA \cite{rana2018cloud}}
		\addplot+[const plot][color=magenta,mark=star,thick] coordinates {
			(1000,79.91) (800,59.92) (600,48.91) (400,36.88) (200,15.92) (100,7.91)
		};\addlegendentry{NSGA-II \cite{guerrero2018multi}}
		\addplot+[const plot][color=cyan,mark=x,thick] coordinates {
			(1000,73.95) (800,58.57) (600,44.63) (400,29.12) (200,14.69) (100,7.67)
		};\addlegendentry{HGAPSO \cite{sharma2016multi}}
		\addplot+[const plot][color=black,mark=oplus,ultra thick] coordinates {
			(1000,59.95) (800,51.76) (600,46.63) (400,30.92) (200,15.49) (100,7.69)
		};\addlegendentry{Proposed WOGA}
		
		\end{axis}
	\end{tikzpicture}}
}\hfill
	\subfigure[Number of active servers ]{%\includegraphics[width=.315\textwidth]{Results/Apms}
		\resizebox{0.3\textwidth}{!}{\begin{tikzpicture}
			\begin{axis}[
			width=0.55\textwidth,
			height=0.45\textwidth,
			%axis lines=left,
			%axis on top=true,
			xmin=100,
			xmax=1000,
			xlabel={Number of VMs},
			xticklabel style={/pgf/number format/1000 sep=},
			ymin=40,
			ymax=650,
			ylabel={Number of active servers},
			legend style={at={(0.5,-0.4)},
			anchor=south,legend columns=3},
			ymajorgrids=true,
			xmajorgrids=true,
			grid style=dashed,
			]
			\addplot[color=Mycolor,mark=triangle,thick] coordinates {
				(1000,621) (800,518) (600,386) (400,220) (200,109) (100,58)
			};\addlegendentry{SMOOP \cite{han2017reducing}}
			
			\addplot [color=purple,mark=triangle,thick] coordinates { 
				(1000,617) (800,443) (600,360) (400,239) (200,118) (100,60)
			};\addlegendentry{Best-Fit \cite{shirvastava2017best}}
			\addplot [color=green,mark=star,thick] coordinates { 
				(1000,609) (800,419) (600,358) (400,240) (200,113) (100,60)};\addlegendentry{Random-Fit \cite{jung2010mistral}}
			\addplot [color=red,mark=o,thick] coordinates { 
				(1000,523) (800,412) (600,315) (400,220) (200,105) (100,52)};\addlegendentry{First-Fit \cite{jangiti2019aggregated}}

			\addplot [color=blue,mark=square,thick] coordinates { 
				(1000,543) (800,423) (600,346) (400,236) (200,115) (100,57)};\addlegendentry{GA based \cite{tseng2017dynamic}}
			\addplot[color=mypink,mark=otimes,thick] coordinates {
				(1000,515) (800,409) (600,310) (400,232) (200,112) (100,51)
			};\addlegendentry{WOA \cite{rana2018cloud}}
			\addplot[color=magenta,mark=star,thick] coordinates {
				(1000,501) (800,417) (600,301) (400,219) (200,100) (100,49)
			};\addlegendentry{NSGA-II \cite{guerrero2018multi}}
			\addplot[color=cyan,mark=x,thick] coordinates {
				(1000,487) (800,396) (600,288) (400,205) (200,110) (100,51)
			};\addlegendentry{HGAPSO \cite{sharma2016multi}}
			
			\addplot[color=black,mark=oplus,ultra thick] coordinates {
				(1000,471) (800,379) (600,282) (400,197) (200,99) (100,46)
			};\addlegendentry{Proposed WOGA}
			
			\end{axis}
			\end{tikzpicture}}
	
}\hfill
	\subfigure[Execution time ]{%\includegraphics[width=.315\textwidth]{Results/exec.time}
	\resizebox{0.3\textwidth}{!}{\begin{tikzpicture}
		\begin{axis}[
		width=0.55\textwidth,
		height=0.45\textwidth,
		%axis lines=left,
		%axis on top=true,
		xmin=180,
		xmax=1000,
		ymin=100,
		ymax=11050,
		ybar stacked,
		bar width=25pt,
		%nodes near coords,
		enlargelimits=0.08,
		legend style={at={(0.5,-0.180)},
			anchor=north,legend columns=3},
		ylabel={Execution Time ($\times 10^{-2}$ sec)},
		xlabel={Number of VMs},
		xticklabel style={/pgf/number format/1000 sep=},
		%symbolic x coords={100, 200, 400, 600,		800, 1000},
		xtick=data,
		ymajorgrids=true,
		xmajorgrids=true,
		grid style=dashed,
		%x tick label style={rotate=0},
		]
		
		\addplot+[ybar,fill=Mycolor,color=Mycolor] plot coordinates {(1000,1439.92) (800,621.45) (600,415.14) (400,117.286) (200,92.09)};\addlegendentry{SMOOP \cite{han2017reducing}}
		
		\addplot+[ybar,fill=black, color=black] plot coordinates {
			(1000,4524.10) (800,2292.85) (600,848.02) (400,441.66) (200,72.816)  
			
		};\addlegendentry{Proposed WOGA}
		\addplot+[ybar,fill=blue,color=blue] plot coordinates {(1000,3436.58) (800,1044.42) (600,166) (400,206) (200,148.18)  };\addlegendentry{GA based \cite{tseng2017dynamic}}
		\addplot+[ybar,fill=mypink,color=mypink] plot coordinates {
			(1000,156) (800,151) (600,483) (400,65) (200,36) 
		};\addlegendentry{WOA \cite{rana2018cloud}}
		\addplot+[ybar,fill=magenta!60!,color=magenta!60!] plot coordinates {
			(1000,876) (800,1099.52) (600,439.59) (400,127.88) (200,175.62)  };\addlegendentry{NSGA-II \cite{guerrero2018multi}}
		\addplot+[ybar,fill=cyan!50!,color=cyan!50!] plot  coordinates { 
			(1000,2116) (800,777) (600,1889) (400,630) (200,438) 
		};\addlegendentry{HGAPSO \cite{sharma2016multi}}
		
		\end{axis}
		\end{tikzpicture}}
}
	
	\caption{Comparison of SM-VMP Framework with State-of-the-arts in Static VMP scenario for varying size of data centre}
	\label{staticVMP}
	
\end{figure*}

The VMs configurations are fixed and pre-defined as stated in Table \ref{table:vm} in case of static VMP.
Fig. \ref{staticVMP} shows comparison of proposed and existing approaches with respect to multiple objectives for static VMP. It is observed from Fig. \ref{staticVMP}(a) that proposed WOGA algorithm allows 62\% to 64\% resource utilization and outperforms HGAPSO, NSGA-II, WOA, SMOOP, GA by 2.1\%, 6.7\%, 5.6\%, 9.8\%, 8.7\% respectively. Moreover, the improvement of 18\%, 20\% and 19\% are observed against first-fit, best-fit and random-fit heuristics, which achieved resource utilization up to 46\%, 44\% and 45\% respectively. It is noticed that secure VMP resists further consolidation of VMs and the resource utilization compromises in order to minimize the conflicting servers. The comparison of side-channel attack probability using proposed security approach is shown in Fig. \ref{staticVMP}(b), where the conflicting servers are reduced up to 25.3\%, 46.9\%, 4.46\%, 8.4\%, 14.5\%, 48.9\%, 57.9\% and 52.4\% over HGAPSO, NSGA-II, WOA, SMOOP, GA, first-fit, best-fit and random-fit based VMP respectively in case of 100 VMs. Fig. \ref{staticVMP}(c) shows comparison of communication cost obtained by using proposed communication-cost model. It is observed that WOGA along with different existing approaches show variable range of results between 10\% and 30\%. However, the average reduction of 25.7\% in communication cost, is achieved by WOGA over existing approaches. The power consumption is shown in Fig. \ref{staticVMP}(d) where proposed WOGA scales down power consumption by 18.93\%, 24.97\%, 25.26\%, 33.69\%, 32.57\%, 29.22\%, 50.10\%, and 26.59\% over HGAPSO, NSGA-II, WOA, SMOOP, GA, first-fit, best-fit and random-fit based VMPs, respectively. Figs. \ref{staticVMP}(e) and \ref{staticVMP}(f) show comparison of the number of active servers and operational execution time consumed during VMP for different size of data centres.  
Furthermore to be noticed that with small size of data centre i.e., 100-200 VMs, the number of active servers and execution time are almost same for existing approaches as well as proposed SM-VMP framework. However, with increase in size of data centre, there is substantial reduction in both number of active servers as well as execution time in case of WOGA which proves its efficacy over the state-of-the-art approaches. {Fig. \ref{fig:pareto-front} compares pareto-front for static VMP of 1000 VMs that depicts the contradictory behavior of four optimization objectives: resource utilization ($RU$), power consumption ($PW$), side-channel attack ($\phi$) and communication cost ($\vartheta$)}.
 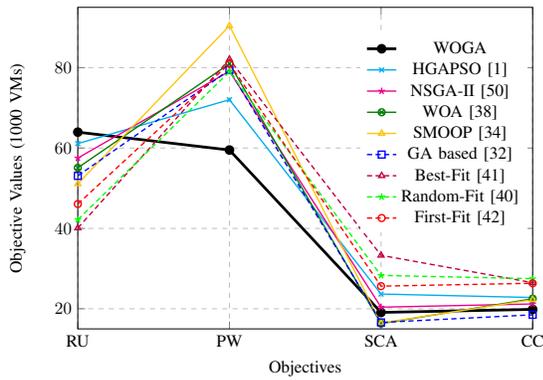
\begin{figure}[htbp!]
 	\centering
 	\resizebox{0.4\textwidth}{!}{\begin{tikzpicture}
 		\begin{axis}[
 		width=0.6\textwidth,
 		height=0.45\textwidth,
 		%axis lines=left,
 		%axis on top=true,
 		xticklabel style={/pgf/number format/1000 sep=},
 		ymin=15,
 		ymax=95,
 		%legend pos=north east,
 		legend style={at={(0.805,0.3)},
 			anchor=south ,draw=none,legend columns=1},
 		ymajorgrids=true,
 		xmajorgrids=true,
 		enlarge x limits=false,
 		grid style=dashed,
 		ylabel={Objective Values (1000 VMs)},
 		xlabel={Objectives},symbolic
 		x coords={RU, PW, SCA, CC},xtick=data]
 		\addplot+[color=black,mark=oplus,ultra thick, sharp plot]
 		plot coordinates {(RU,63.95) (PW,59.5) (SCA,19.10) (CC,19.85)};\addlegendentry{ WOGA}
 		\addplot+[color=cyan,mark=x,thick,sharp plot] plot coordinates {(RU,61.10) (PW,72.02) (SCA,23.67) (CC,22.8)
 		};\addlegendentry{HGAPSO \cite{sharma2016multi}}
 		
 		\addplot+[color=magenta,mark=star,thick,sharp plot] plot coordinates {(RU,57.5) (PW,79.1) (SCA,20.4) (CC,21.2)
 		};\addlegendentry{NSGA-II \cite{guerrero2018multi}}

 		\addplot+[color=mypink,mark=otimes,thick,sharp plot]plot coordinates {(RU,55.07) (PW,80.98) (SCA,16.3) (CC,22.5)
 		};\addlegendentry{WOA \cite{rana2018cloud}}

 		\addplot+[color=Mycolor,mark=triangle,thick,sharp plot] plot coordinates {(RU,51.09) (PW,90.42) (SCA,16.47) (CC,22.264)
 		};\addlegendentry{SMOOP \cite{han2017reducing}}
 		
 		\addplot+[color=blue,mark=square,thick,sharp plot] plot coordinates {(RU,53.07) (PW,79.428) (SCA,16.6) (CC,18.5)};\addlegendentry{ GA based \cite{tseng2017dynamic}}
 		
 		\addplot+[color=purple,mark=triangle,thick,sharp plot] plot coordinates {(RU,40.1) (PW,82.08) (SCA,33.3) (CC,26.5)};\addlegendentry{Best-Fit \cite{shirvastava2017best}}
 		
 		\addplot+[color=green,mark=star,thick,sharp plot] plot coordinates {(RU,42.17) (PW,78.98) (SCA,28.3) (CC,27.5)};\addlegendentry{Random-Fit\cite{jung2010mistral} }
 		
 		\addplot+[color=red,mark=o,thick,sharp plot] plot coordinates {(RU,46.07) (PW,80.98) (SCA,25.63) (CC,26.35)};\addlegendentry{First-Fit \cite{jangiti2019aggregated}}
 		
 		\end{axis}
 		\end{tikzpicture}}
 	\caption{Pareto-front: WOGA vs State-of-the-arts }
 	\label{fig:pareto-front}
 \end{figure}

\subsubsection{Dynamic VMP}
Dynamic VMP allows allocation of VMs whose configuration is defined at run-time and handles de-allocation and re-allocation of variable size VMs dynamically by migrating VMs from over/under-utilized servers to the selected near-optimal server. {To compare the performance of SM-VMP Framework against state-of-the-arts for dynamic VMP, the CPU and memory utilization percent of VMs are extracted from the benchmark Google Cluster Dataset (GCD) \cite{reiss2011google} recorded at time-interval of five minutes}. Here, the number of users are taken as 20 percent of the size of the data centre (though it can be changed as per the requirement). In the observed simulation, various performance metrics including $RU$, $PW$, $\phi$, $\vartheta$, active physical machines (APMs), number of migrations ($MG$\#), migration cost ($MG_{cost}$) and  execution time ($Time$) are analysed and comparative results are shown in Table \ref{table:dynamicVMP}.

\begin{table*}[htbp]
	
	\caption[Table caption text] {Proposed vs State-of-the-arts comparison for Dynamic VMP }  %\cite[p.10]{refid} }
	\label{table:dynamicVMP}
	\small
	%\centering 
	%\resizebox{0.8\textwidth}{!}{\begin{minipage}{\textwidth}
	%\resizebox{12cm}{!}{
		\centering
		\begin{tabular}{ |l|l|l|cccccccc| }
			\hline
			& & &\multicolumn{8}{c|}{Performance metrics} \\
			\cline{4-11}
			Approach &  VMs & Users&  RU (\%) & PW (W) &$\phi$&$\vartheta$&APMs&$MG_{cost}$& $MG$\# 
			 & $Time$ \\ \hline
			
			\multirow{3}{*}{WOGA} & 100& 20&68.21&2.53E+3&16.19&23.91&19 &1.35E+1& 18&5.73   \\
			& 400&80 & 68.43& 1.03E+4&18.45 & 14.72& 65& 3.33E+2& 45 &32.61\\
			& 800&160 &67.67&  1.84E+4&17.9 & 18.13& 127& 1.57E+3&76  &108.61\\
			& 1200&240 &69.35&2.91E+4&21.2&27.84&216& 9.61E+2&93 &196.78\\
			
			\hline
				
			\multirow{3}{*}{HGAPSO \cite{sharma2016multi}} & 100&20 &60.10&3.30E+3&19.82&22.14 &20&9.05E+1 &20 &16.37  \\
			& 400&80 & 56.05& 1.26E+4& 21.41& 15.59& 80& 5.47E+2& 49 &154.62\\
			& 800&160 &59.50& 2.69E+4& 20.94& 25.88& 157& 2.81E+3&106  &678.58\\
			& 1200&240 &55.22&3.60E+4&19.38&26.96&236& 5.08E+3&95  &1138.30\\
			
			\hline

			\multirow{3}{*}{NSGA-II \cite{guerrero2018multi}}  & 100&20 &64.98&3.12E+3&13.7&11.78&20 &7.81E+1& 49 & 8.12  \\
			& 400&80 & 64.57& 1.96E+4& 15.8& 19.8& 80& 1.69E+2&201  &78.92\\
			& 800& 160&63.34& 3.31E+4& 21.9& 15.34& 158& 1.20E+3& 354 &453.56\\
			& 1200&240 &63.87&3.59E+4&24.7&23.62&236& 3.06E+3& 512 &679.16\\
			
			\hline
				\multirow{3}{*}{GA \cite{tseng2017dynamic}} & 100&20 &56.37&3.35E+3&14.9&33.3&20 &4.69E+2&30  &7.46  \\
			& 400&80 &58.64 & 1.39E+4& 33.0& 34.4& 80& 1.84E+3& 112 &134.65\\
			& 800&160 &54.46& 2.69E+4& 25.8& 32.7& 158& 4.29E+3&478  &217.90\\
			& 1200&240 & 57.83&3.40E+4&31.42&30.47&212& 6.64E+3&513  &421.34\\
			
			\hline
			
				\multirow{3}{*}{SMOOP\cite{han2017reducing}} & 100& 20&56.71&3.41E+3&10.52&16.6&23 &4.74E+1& 57 &4.92   \\
			& 400&80 & 55.19& 1.52E+4& 9.61& 19.2&87 &9.18E+2& 119 &25.63\\
			& 800&160 & 57.61& 4.01E+4& 11.82& 22.15&164 &10.59E+3& 402  &121.81\\
			& 1200& 240&55.43&4.54E+4&12.91&20.19&237& 1.89E+4& 594 &164.50\\
			
			\hline
				
			\multirow{3}{*}{WOA \cite{rana2018cloud}} & 100& 20&59.16&3.20E+3&13.125&9.52&20 &4.8E+1& 33 &5.81   \\
			& 400&80 & 56.71& 1.35E+4& 20.5& 15.35&80&5.40E+2   &184 &37.43\\
			& 800&160 &56.84& 2.71E+4& 18.19& 19.31& 160& 2.37E+3&324  &129.55\\
			& 1200&240 &55.91&3.32E+4&19.25&18.18& 239&2.94E+3& 416  &162.20\\
			
			\hline
				
			\multirow{3}{*}{First-Fit \cite{jangiti2019aggregated}} & 100&20 &63.91&2.68E+3&45.0&38.9&14 &6.50E+1& 14 &1.32  \\
			& 400&80 & 64.01& 1.06E+4& 42.5& 44.2& 69& 8.95E+2& 49  &9.17\\
			& 800&160 &63.63& 2.16E+4& 31.25& 46.6& 116& 2.20E+2& 82  &38.29\\
			& 1200&240 &64.69&2.18E+4&44.5&39.3&226& 1.89E+2& 119 &98.19\\
			
			\hline
				
			\multirow{3}{*}{Best-fit \cite{shirvastava2017best}} & 100&20 &44.56&2.24E+3&32.1&34.8&20 &3.4E+2& 26  &3.67  \\
			& 400&80 & 47.82& 1.11E+4& 37.4& 42.1& 80& 2.2E+3& 69 &25.89\\
			& 800& 160&46.23& 2.42E+4& 40.9&35.6 & 160& 2.50E+3& 137  &89.23\\
			& 1200&240 &49.84&3.01E+4&36.3&42.5&239& 2.37E+3& 209 &148.7\\
			
			\hline
				
			\multirow{3}{*}{Random-Fit \cite{jung2010mistral}} & 100&20 &36.7&2.63E+3&35.4&47.7&20 &2.98E+2& 18  &2.22  \\
			& 400&80 & 37.84& 1.12E+4& 37.5& 49.3& 80& 2.09E+3& 79  &18.34\\
			& 800&160 &39.25& 2.24E+4& 35.6& 42.4& 158& 1.42E+3& 307 &59.47\\
			& 1200&240 &39.14&2.81E+4&33.1&44.9& 238&1.62E+3 & 409 &103.27\\
			
			\hline

	\end{tabular}
\end{table*}

The simulation results show that proposed WOGA outperforms all the comparative existing schemes in resource utilization and power consumption for each size of data centre. The maximum resource utilization observed is 69.35\% for data centre of 1200 VMs which surpasses 55.22\% to 64.69\% resource utilization produced by the existing VMPs. The maximum improvement observed in resource utilization is up to 30.21\% over random-fit. Though power consumption increases with size of data centre, WOGA shows least consumption of power each time as compared to existing approaches. Furthermore, it is observed that network traffic/communication cost and probability of co-residency attacks are dependent on number of active users and independent of size of data centre. With increase in number of users, probability of security threats scales up and users having large number of data-intensive VMs (having memory $\ge$ CPU in reported experiments) raises communication-cost. The approach in proposed SM-VMP framework shows equal or lesser security threat probability as compared to the above mentioned existing schemes, except SMOOP  \cite{han2017reducing} and GA+Weighted Sum approach \cite{tseng2017dynamic} because here, the pareto-front is computed with maximum weightage i.e., 40\% given to security feature over other performance metrics. Moreover, WOGA reduces security threats by 28.81\% against first-fit. 
 As we have implemented the communication cost by using proposed formulation of $\vartheta$ for all the existing comparative approaches, there is not much variation and communication cost lies in the range of (9\% to 30\%). However, SM-VMP reduces communication cost up to 23\%, 20.5\% and 25.7\% over first-fit, best-fit and random-fit VMPs respectively. Additionally, the number of active servers, incurred number of migrations and migration cost (while handling de-allocation/re-allocation and under/over-utilized servers) are also reduced. The operational execution time for first-fit, best-fit and random-fit is least as they do not require iterative optimization, while the existing approaches with weighted sum pareto-front (which is biased towards dominant objective) viz. SMOOP and GA based VMP show lesser execution time as compared to proposed WOGA, for small size of data centre. However, WOGA  beats all other existing approaches in terms of execution time with maximum reduction up to 82.21\% over HGAPSO.

\begin{figure*}[htbp]
	
	\centering
	
	\subfigure[Power and Resource utilization state transitions vs number of epochs ]{\includegraphics[width=.325\textwidth]{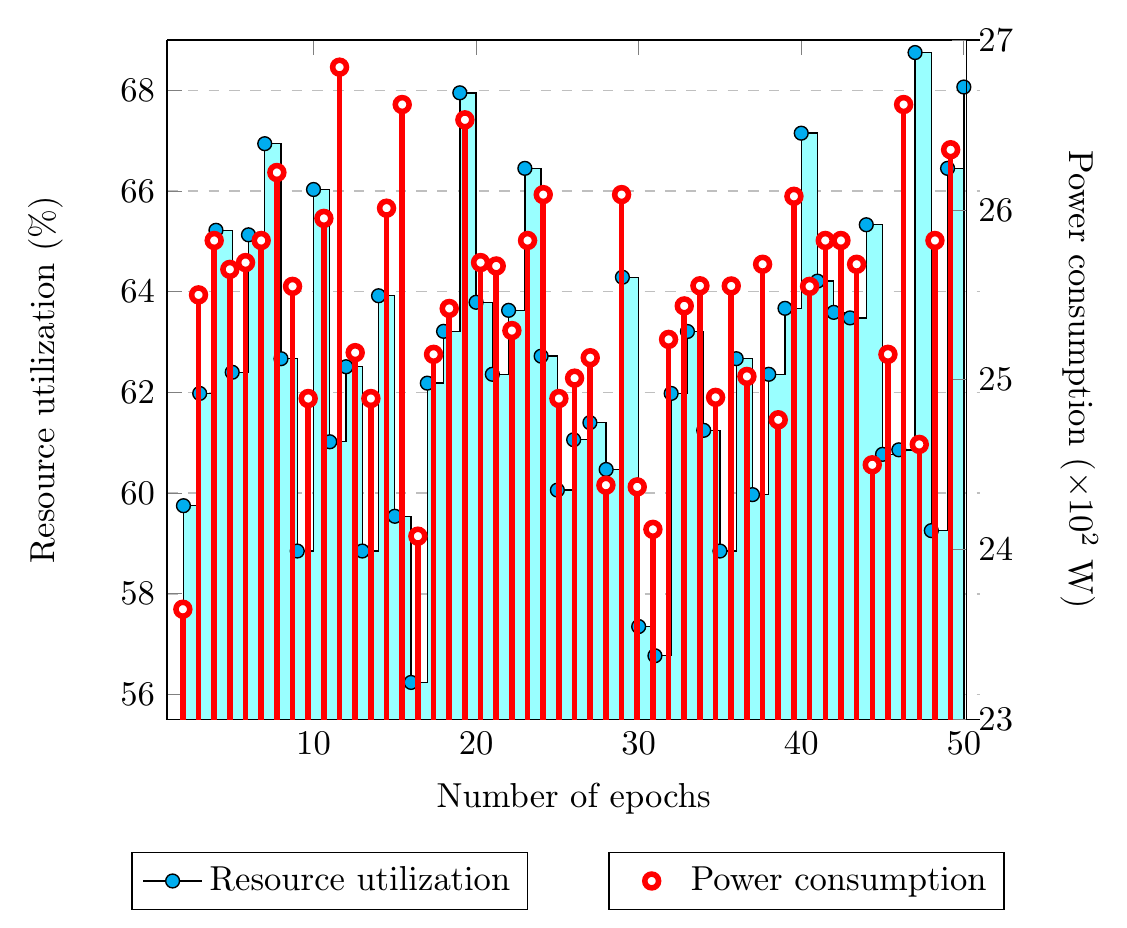}}\hfill
	\subfigure[Co-residency and Communication cost state transitions vs number of epochs  ]{\includegraphics[width=.325\textwidth]{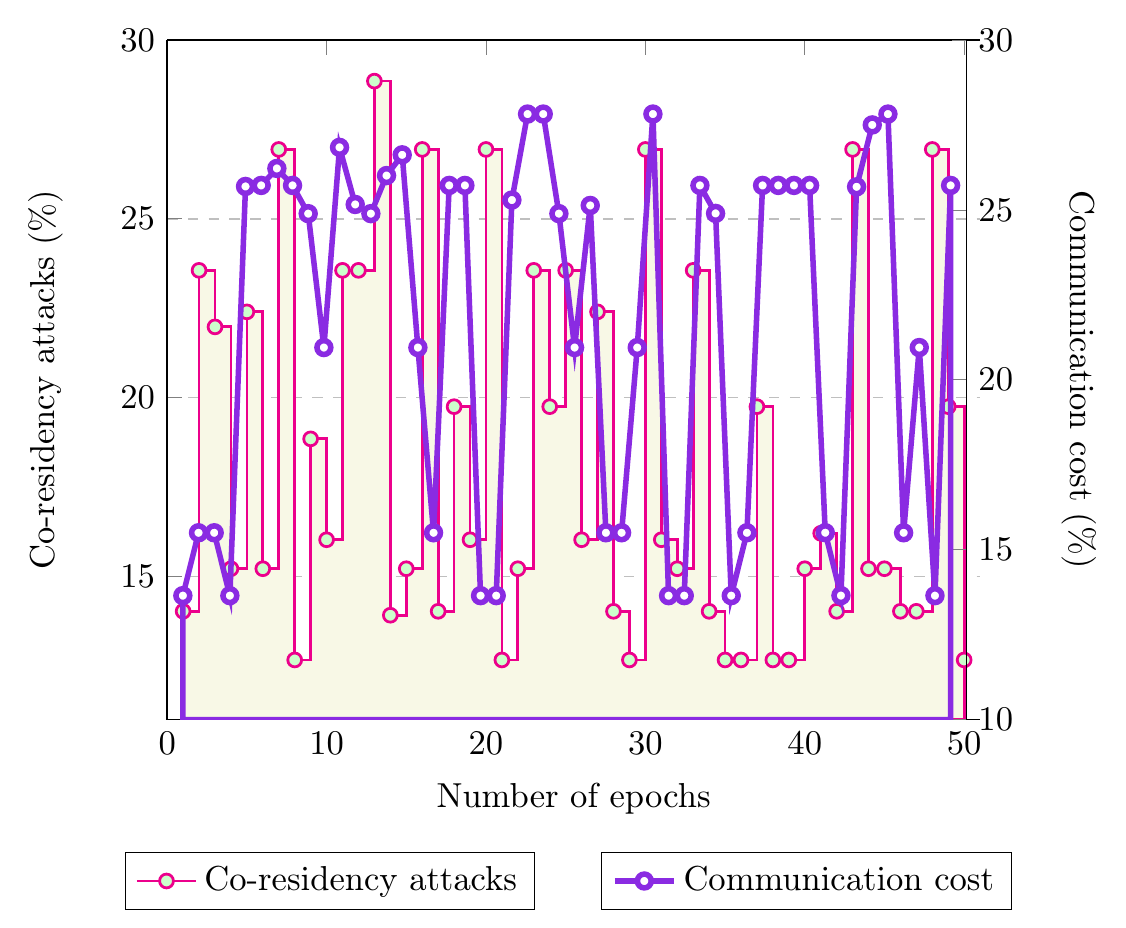}}\hfill
	\subfigure[Boxplots of multiple objectives  ]{\includegraphics[width=.295\textwidth]{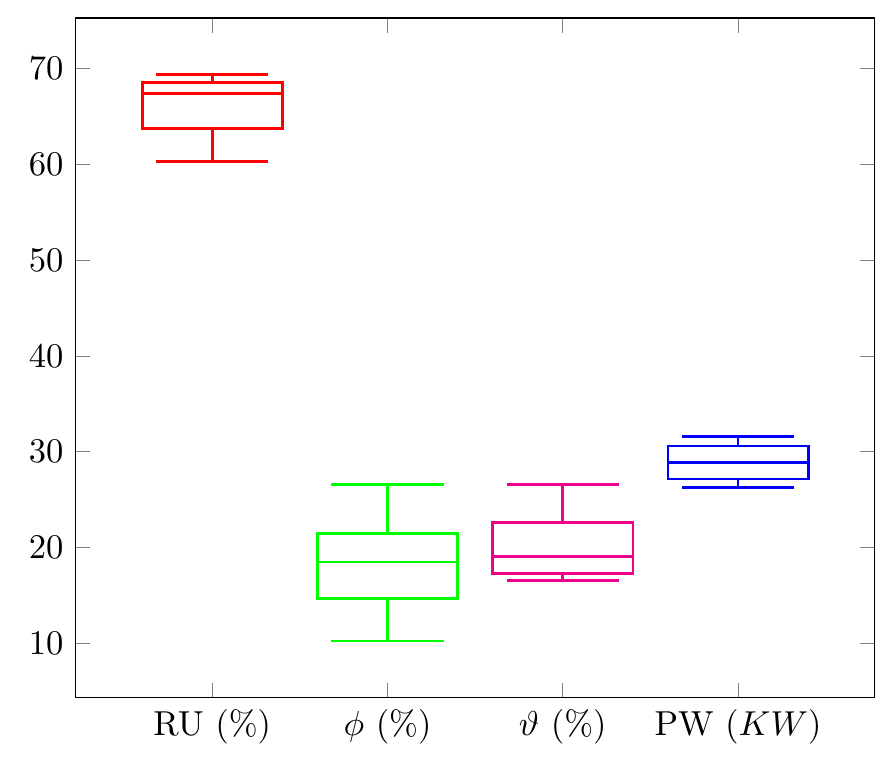}}
	
	\caption{Performance metrics for Dynamic VMP }
	\label{fig:dynamicVMP}
	
\end{figure*}

Fig. \ref{fig:dynamicVMP}(a) shows power consumption and resource utilization state-transitions over the number of epochs, as observed during proposed WOGA based dynamic VMP of 100 VMs. On the same lines, probability of co-residency attacks and communication cost transitions versus number of epochs are shown in Fig. \ref{fig:dynamicVMP}(b). The boxplots shown in Fig. \ref{fig:dynamicVMP}(c) distributes performance metrics viz. RU, $\phi$, $\vartheta$ and power consumption statistically through quartiles. The bottom, middle and top of the box are the first, second (i.e., the median) and third quartiles. The ends of the whiskers are the minimum
and maximum of all values, respectively, achieved for dynamic VMP run of 1200 VMs.

\section{Conclusions and Future Work}
 A novel framework is proposed to provide a pareto-optimal solution for VM consolidation problem entangled with multiple issues to concurrently serve the perspectives of both the cloud user and service provider.
Furthermore, a multi-efficient VM migration scheme is introduced to re-optimize VMs allocation during dynamic VMP. There is substantial 
reduction in security attacks, inter-communication cost and power consumption, and improvement in resource utilization during VM placement. Evidently, the results show superiority of SM-VMP framework compared to the existing state-of-the-art approaches. In future, the proposed VM placement strategy can be extended by prioritizing the objectives as per the dynamic requirement, adding objectives like trust and reliability based VM allocation scheme. Also, {the proposed VMP approach can be extended to incorporate an online VM mapping across the clusters}.

%\ifCLASSOPTIONcompsoc
 % \section*{Acknowledgments}
%\else
%\fi

%The author would like to thank National Institute of Technology, Kurukshetra, India for financially supporting the research work.

%\ifCLASSOPTIONcaptionsoff
%  \newpage
%\fi

%\nocite{*} 

\bibliographystyle{IEEEtran}
\bibliography{mynewbibfile}

\begin{IEEEbiography}[{\includegraphics[width=0.7\linewidth]{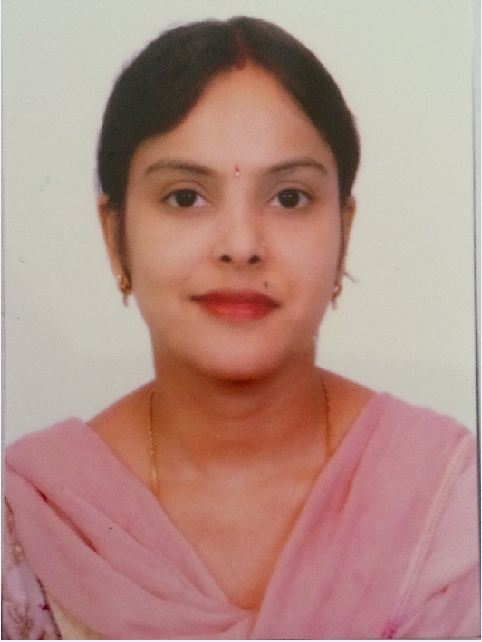}}]{Deepika Saxena}
	received M.Tech degree in Computer Science and Engineering  from Kurukshetra University Kurukshetra, India. Currently, she is pursuing Ph.D from Department of Computer Applications, National Institute of Technology, Kurukshetra, India. Her major research interests are Machine Learning, Evolutionary Algorithms, Resource Management, and Security in Cloud Computing.
\end{IEEEbiography}
\vskip 0pt plus -1fil 
\begin{IEEEbiography}[{\includegraphics[width=0.7\linewidth]{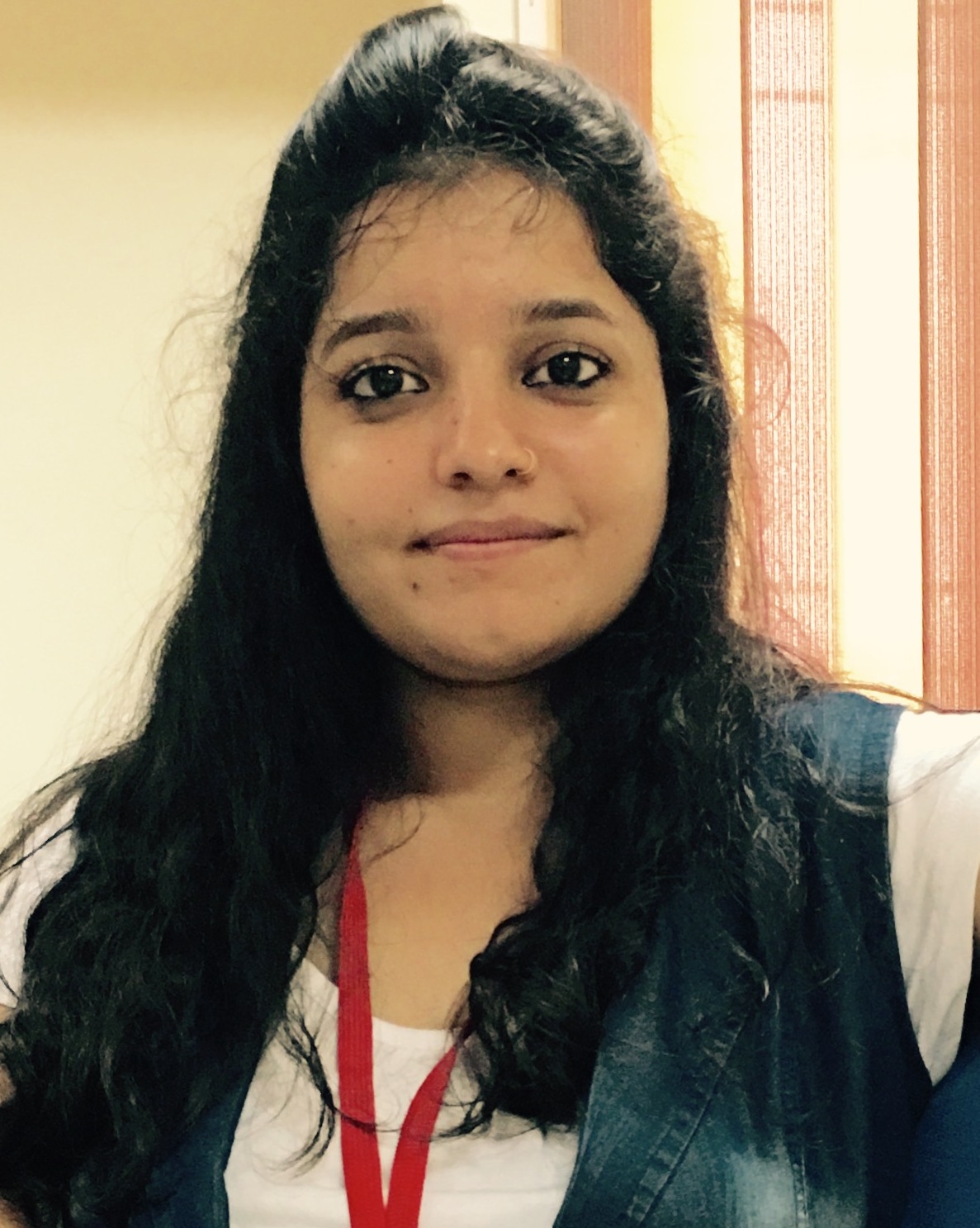}}]{Ishu Gupta} received MCA (Gold Medalist) from Kurukshetra University, Kurukshetra, India, in 2015. She earned her Ph.D. and working as a research assistant with the Department of Computer Applications, National Institute of Technology, Kurukshetra, India. Her major research interests are cloud computing, machine learning, information security \& privacy. She has more than 25 publications and was the recipient of the excellent paper award twice.
\end{IEEEbiography}
\vskip 0pt plus -1fil 
\begin{IEEEbiography}[{\includegraphics[width=0.7\linewidth]{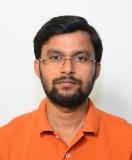}}]{Jitendra Kumar} is an Assistant Professor with the Department of Computer Applications, National Institute of Technology Tiruchirappalli, India. He earned his doctorate from the National Institute of Technology Kurukshetra, India in 2019. His current research interests include Cloud Computing, Machine Learning, Data Analytics, Parallel Processing. He has published more than 25 articles in international journals and conferences of repute.
\end{IEEEbiography}
\vskip 0pt plus -1fil 
\begin{IEEEbiography}[{\includegraphics[width=0.7\linewidth]{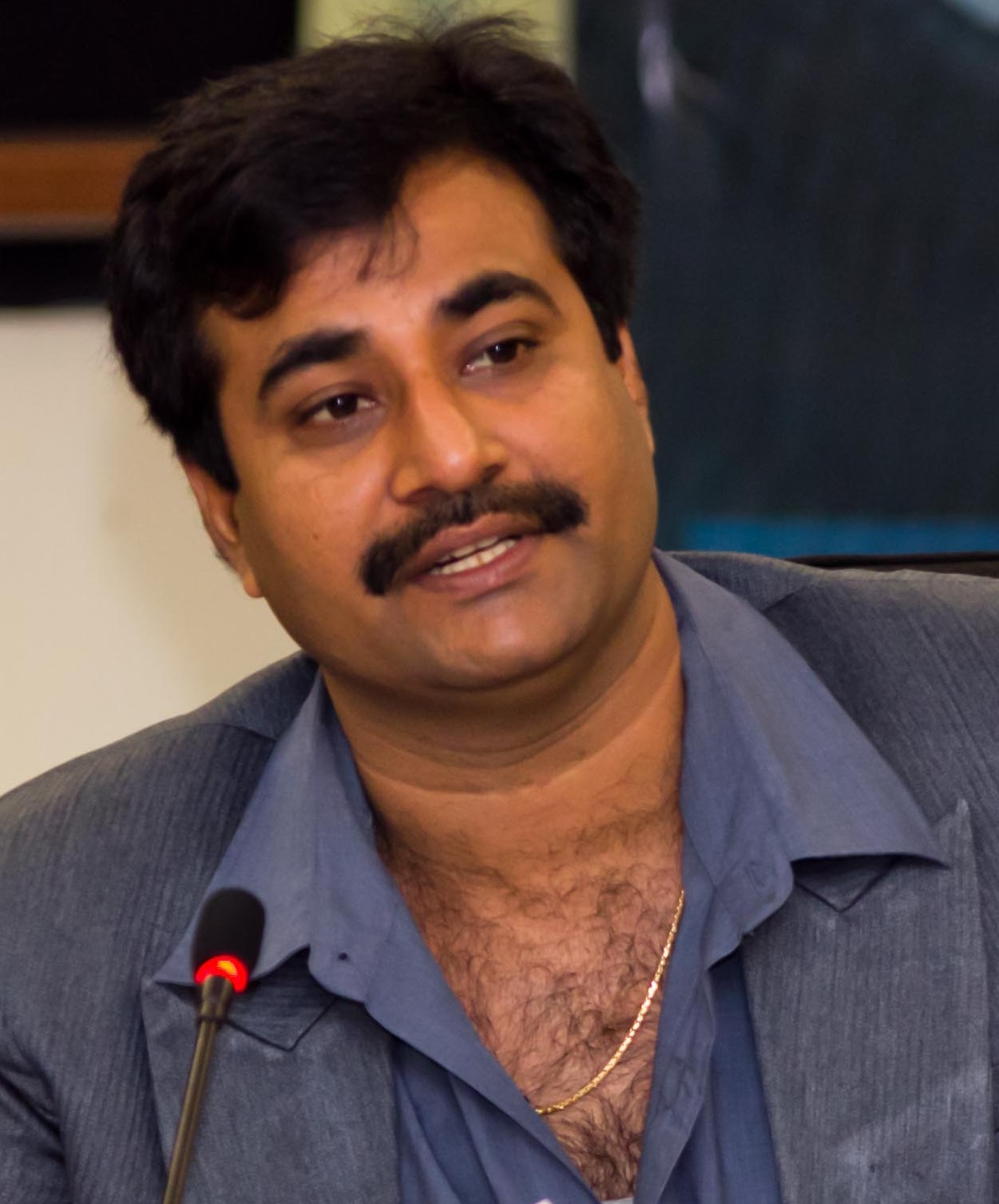}}]{Ashutosh Kumar Singh}
	is working as a Professor and Head in the Department of Computer Applications, National Institute of Technology Kurukshetra, India. He received his PhD in Electronics Engineering from Indian Institute of Technology, BHU, India and Post Doc from Department of Computer Science, University of Bristol, UK. His research area includes Design and Testing of Digital Circuits, Data Science, Cloud Computing, Machine Learning, Security. He has published more than 300 research papers in different journals of high repute.
	
\end{IEEEbiography}
\vskip 0pt plus -1fil 
\begin{IEEEbiography}[{\includegraphics[width=0.7\linewidth]{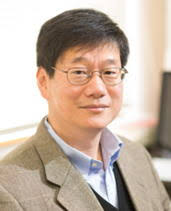}}]{Xiaoqing Wen} received the Ph.D. degree from Osaka University, Japan, in 1993. He founded Dependable Integrated Systems Research Center in 2015 and served as its Director until 2017. His research interests include VLSI test, diagnosis, and testable design. He holds 43 U.S. Patents and 14 Japan Patents on VLSI testing. He is a fellow of the IEEE, a senior member of the IPSJ, and a senior member of the IEICE. He is serving as associate editors IEEE Transactions on VLSI and the Journal of Electronic Testing: Theory and Applications.
	
\end{IEEEbiography}

\end{document}